\documentclass[11pt]{article}
\pdfoutput=1 
\usepackage{amssymb}
\usepackage{amsmath}
\usepackage{amstext}
\usepackage{graphicx,epsfig}
\usepackage{epsfig}
\usepackage{verbatim} 
\usepackage{fancybox}
\usepackage{color}
\usepackage{ulem}
\usepackage{enumitem}
\usepackage{youngtab}
\usepackage{subfigure}
\usepackage{bbm}
\usepackage{parskip}
\usepackage[numbers,sort&compress]{natbib}
\usepackage{ytableau}
\usepackage[utf8]{inputenc}

\usepackage{tikz}
\usetikzlibrary{decorations}
\pgfdeclaredecoration{complete sines}{initial}
{
    \state{initial}[
        width=+0pt,
        next state=upsine,
        persistent precomputation={\pgfmathsetmacro\matchinglength{
            \pgfdecoratedinputsegmentlength / int(\pgfdecoratedinputsegmentlength/\pgfdecorationsegmentlength)}
            \setlength{\pgfdecorationsegmentlength}{\matchinglength pt}
        }] {}
    \state{upsine}[width=\pgfdecorationsegmentlength,next state=downsine]{
        \pgfpathsine{\pgfpoint{0.25\pgfdecorationsegmentlength}{0.5\pgfdecorationsegmentamplitude}}
        \pgfpathcosine{\pgfpoint{0.25\pgfdecorationsegmentlength}{-0.5\pgfdecorationsegmentamplitude}}
    }
    \state{downsine}[width=\pgfdecorationsegmentlength,next state=upsine]{
        \pgfpathsine{\pgfpoint{0.25\pgfdecorationsegmentlength}{-0.5\pgfdecorationsegmentamplitude}}
        \pgfpathcosine{\pgfpoint{0.25\pgfdecorationsegmentlength}{0.5\pgfdecorationsegmentamplitude}}
}
    \state{final}{}
}

\def\stu{{St\"uckelberg }}

\newcommand{\Comment}[1]{{}}
\definecolor{darkblue}{rgb}{0.15,0.35,0.55}
\definecolor{reddish}{rgb}{0.65, 0.2, 0.2}
\usepackage[linktocpage=true]{hyperref}
\hypersetup{
colorlinks=true,
citecolor=darkblue,
linkcolor=reddish,
urlcolor=darkblue,
pdfauthor={},
pdftitle={},
pdfsubject={}
}

\definecolor{green3}{RGB}{44, 160, 44}

\newcommand{\be}{\begin{equation}}
\newcommand{\ee}{\end{equation}}
\newcommand{\bea}{\begin{eqnarray}}
\newcommand{\eea}{\end{eqnarray}}
\newcommand{\beas}{\begin{eqnarray*}}
\newcommand{\eeas}{\end{eqnarray*}}
\newcommand{\nn}{\nonumber}

\def\mn{_{\mu \nu}}

\definecolor{darkred}{rgb}{0.7,0.3,0.3}
\definecolor{darkgreen}{rgb}{0.2,0.7,0.3}
\definecolor{lightgreen}{rgb}{.816,.94,.753}
\definecolor{greyish}{rgb}{.8,.8,.8}
\definecolor{darkblue2}{rgb}{0.3,0.4,0.9}
\usepackage{pifont}
\usepackage{xcolor,colortbl}

\def\({\left(}
\def\){\right)}

\newcommand{\bm}{{\gamma}}

\newcommand{\half}{\frac{1}{2}}

\def\gsim{ \lower .75ex \hbox{$\sim$} \llap{\raise .27ex \hbox{$>$}} }
\def\lsim{ \lower .75ex \hbox{$\sim$} \llap{\raise .27ex \hbox{$<$}} }

\def\xyma{\xymatrix@M.7em}
\def\xymas{\xymatrix@M.1em}

\title{}
\author{}

\numberwithin{equation}{section}

\begin{document}
%
\renewcommand{\thefootnote}{\fnsymbol{footnote}}
~
\vspace{1.75truecm}
\begin{center}
{\huge \bf{Shift-Symmetric Spin-1 Theories}}\\ \vspace{.2cm}
{\LARGE \bf{}}
\end{center} 

\vspace{1truecm}
\thispagestyle{empty}
\centerline{\Large James Bonifacio,${}^{\rm a,}$\footnote{\href{mailto:james.bonifacio@case.edu}{\texttt{james.bonifacio@case.edu}}} Kurt Hinterbichler,${}^{\rm a,}$\footnote{\href{mailto:kurt.hinterbichler@case.edu} {\texttt{kurt.hinterbichler@case.edu}}} Laura A. Johnson,${}^{\rm a,}$\footnote{\href{mailto:lxj154@case.edu} {\texttt{lxj154@case.edu}}} and Austin Joyce${}^{\rm b,}$\footnote{\href{mailto:austin.joyce@columbia.edu}{\texttt{austin.joyce@columbia.edu}}} }
\vspace{.5cm}

\centerline{{\it ${}^{\rm a}$CERCA, Department of Physics,}}
 \centerline{{\it Case Western Reserve University, 10900 Euclid Ave, Cleveland, OH 44106}} 
 \vspace{.25cm}
 
 \centerline{{\it ${}^{\rm b}$Center for Theoretical Physics, Department of Physics,}}
 \centerline{{\it Columbia University, New York, NY 10027}} 
 \vspace{.25cm}

 \vspace{.8cm}
\begin{abstract}
\noindent
We study interacting massive spin-1 theories in de Sitter (dS) and anti-de Sitter (AdS) space that possess shift symmetries parametrized by (A)dS Killing vectors. We show how they emerge from the massless limit of massive spin-2 theories on (A)dS space. In the case of massive gravity, the corresponding spin-1 theory realizes a symmetry breaking pattern that takes two copies of the (A)dS isometry group down to a diagonal subgroup. By taking the flat space limit of this theory, we find a new symmetry of the decoupling limit of massive gravity in flat space. This symmetry acts on the vector modes, is parametrized by an antisymmetric tensor, and fixes the nonlinear structure of the scalar-vector sector of the decoupling limit.

\end{abstract}

\newpage

\setcounter{tocdepth}{2}
\tableofcontents
\newpage
\renewcommand*{\thefootnote}{\arabic{footnote}}
\setcounter{footnote}{0}

\section{Introduction and summary}
Shift symmetries for scalar fields provide a powerful organizing structure for a variety of effective field theories. In theories with spontaneously broken internal symmetries, there is a massless scalar Nambu--Goldstone boson for each broken generator of the symmetry group. The broken symmetries act to leading order in fields as shift symmetries, which  protects the masslessness of Goldstone bosons.
From the $S$-matrix point of view, shift symmetries imply that amplitudes vanish as the momentum of an external Goldstone line goes to zero, known as an Adler zero~\cite{Adler:1964um,Adler:1965ga}. Higher-order shift symmetries, where the scalars shift by powers of the spacetime coordinates, are present in galileon theories and their generalizations~\cite{Nicolis:2008in, Trodden:2011xh,Hinterbichler:2014cwa,Griffin:2014bta}.  These can be thought of as Goldstone bosons from the spontaneous breaking of particular spacetime symmetries~\cite{deRham:2010eu,Hinterbichler:2010xn,Goon:2011qf,Goon:2011uw,Goon:2012dy}. These lead to further enhanced soft limits of scattering amplitudes, where the amplitudes go to zero with higher powers of the external momenta \cite{Cheung:2014dqa}.  In exceptional cases these can be used to bootstrap the theories~\cite{Cheung:2015ota, Cheung:2016drk}.

A natural question is whether shift symmetries and enhanced soft limits can be present for fields with non-zero spin.  Independent of whether these fields have an interpretation as Goldstone bosons of some symmetry breaking, we can ask if effective field theories describing spinning particles invariant under shift-like symmetries exist and whether they possess, or are determined by, enhanced soft limits that generalize the Adler zeros.  

Effective field theories for a massless vector field in flat space do not exhibit enhanced single soft limits~\cite{Cheung:2018oki}. 
Similar results were obtained by studying the algebras that such shift symmetries would imply~\cite{Klein:2017npd, Klein:2018ylk,Bogers:2018zeg,Roest:2019oiw,Roest:2019dxy}.  However, in de Sitter (dS) and anti-de Sitter (AdS) space there is a richer variety of possible single-field theories with shift symmetries~\cite{Bonifacio:2018zex}.  Unlike flat space, where only massless fields have coordinate-dependent shift symmetries, massive fields in (A)dS space acquire such symmetries at certain fixed values of their mass relative to the (A)dS scale.  In particular, in (A)dS$_4$ the free massive spin-1 vector field $A_\mu$ acquires shift symmetries at the discrete mass values
\be
m^2 = \frac{(k+2)(k+3)}{L^2} = - H^2 (k+2)(k+3)  \,, \quad k=0,1,2,\ldots, \,
\ee
where $L$ is the AdS radius and $H$ is the Hubble scale in the dS case.
For $k=0$,  the corresponding symmetry acts to shift the field by an (A)dS Killing vector $\xi_\mu$,
\be 
 \delta A_{\mu}=\xi_\mu.
\ee
This is the curved space analogue of the simplest possible global shift symmetry for a vector field.
As discussed in Ref.~\cite{deRham:2018svs}, this vector serves as the longitudinal mode of a massive spin-2 particle in (A)dS space in a particular limit. As we send the mass of a massive spin-2 field to zero while keeping the (A)dS radius fixed, the field decomposes into a massless spin-2 field and a $k=0$ massive vector.

In this paper, we study interacting theories of these $k=0$ vectors.  By studying the algebra of symmetries, we find that there is a unique way to deform the algebra of the free theory.  There are thus two types of shift symmetries, based on whether or not the symmetry algebra of the free theory gets deformed, which we call ``non-abelian'' or ``abelian,'' respectively. Theories that are invariant under these symmetries can be constructed by considering the (A)dS decoupling limit of interacting theories of massive spin-2 particles in (A)dS space.  
In the case of interactions for a massive spin-2 particle with a linear kinetic term, we get an interacting spin-1 theory with an abelian shift symmetry, and in the case of interactions built from the Einstein--Hilbert kinetic term, we get an interacting spin-1 theory with a non-abelian shift symmetry.  If we choose the interactions to be those of the pseudo-linear spin-2 theory extended to (A)dS space \cite{Folkerts:2011ev, Hinterbichler:2013eza,Akagi:2014iea} or ghost-free de Rham--Gabadadze--Tolley (dRGT) massive gravity~\cite{deRham:2010kj}, we get ghost-free spin-1 theories with abelian or non-abelian shift symmetries, respectively.  The ghost-free spin-1 theory with a non-abelian shift symmetry was derived in Ref.~\cite{deRham:2018svs} and here we find its nonlinear symmetries.

The full symmetry algebra of the interacting spin-1 theory with the non-abelian shift symmetry is the direct sum of two copies of the (A)dS isometry algebra,\footnote{Here and throughout we refer to complexified Lie algebras. The relevant real form depends on the spacetime signature and whether we are in dS or AdS space.} $\mathfrak{so}(5)\oplus \mathfrak{so}(5)$.  The actual (A)dS isometry algebra is a diagonal $\mathfrak{so}(5)$ subalgebra, so the symmetry breaking pattern is 
\be \mathfrak{so}(5)\oplus \mathfrak{so}(5)\rightarrow \mathfrak{so}(5)_{\rm diag}.\ee  
The broken generators can be seen as arising from the global part of the \stu diffeomorphism symmetry that survives the decoupling limit. 

In the flat space limit, the (A)dS symmetries reduce to Poincar\'e symmetries, and so the broken generators can be organized into broken translations and broken Lorentz transformations. 
The flat space decoupling limit of massive gravity is described by an interacting
massless scalar-vector-tensor theory where there is a galileon symmetry that acts on the scalar \cite{deRham:2010ik},
\be \label{eq:galshift}
\phi\mapsto \phi+ c+b_\mu x^\mu.
\ee
Here $c$ is a constant, $b_{\mu}$ is a constant vector, and $x^{\mu}$ is the spacetime coordinate.
These galileon symmetries can be thought of as resulting from the global \stu translations that correspond to broken translations.  
However, the action of the broken Lorentz transformations has been unaccounted for until now.  We will see that the broken Lorentz transformations are indeed a symmetry of the scalar-vector sector of the flat space massive gravity decoupling limit action.  In the case of dRGT massive gravity, these new symmetries take the form
 \be  \delta  \hat{A}_\mu=m_{\mu\nu}\left(x^\nu-{2\over \Lambda_3^3} \partial^\nu\hat \phi\right)\, ,\label{fullsymmfinaleeintro}\ee
where $m_{\mu\nu}$ is a constant antisymmetric tensor parametrizing the broken Lorentz transformations and $\Lambda_3 \equiv (M_{\rm Pl}m^2)^{1/3}$ is the strong coupling scale.  This symmetry fixes the form of the complicated scalar-vector interactions in the decoupling limit relative to a few leading terms.

The rest of the paper is organized as follows: in Section~\ref{sec:freetheory} we describe the $k=0$ shift symmetry of a free vector field with mass squared $m^2 = 6L^{-2}$ and show how it arises in the decoupling limit of a linearized massive spin-2 field on (A)dS space. In Section~\ref{sec:algebra} we describe the algebra of symmetries enjoyed by this theory and find a unique deformation such that the shift symmetries no longer commute. In Sections~\ref{sec:abelian} and~\ref{nonlinemgsec} we construct theories that realize the abelian and non-abelian symmetry algebras, respectively. In Section~\ref{sec:flatspace} we describe how the non-abelian shift symmetry of the (A)dS decoupling limit of dRGT massive gravity leads to the non-linear vector symmetry \eqref{fullsymmfinaleeintro} in the flat space decoupling limit of massive gravity. In Section~\ref{sec:concl} we summarize our conclusions and comment on interesting future directions to pursue. Several technical results are collected in the appendices: in Appendix~\ref{projapp} we describe how to construct the non-abelian $k=0$ vector shift symmetry using embedding space techniques. In Appendix~\ref{app:decouplingvector} we describe how the scalar-vector sector of nonlinear massive gravity is fixed by demanding invariance under the symmetry~\eqref{fullsymmfinaleeintro}. Finally, in Appendix~\ref{app:higherkvects} we describe scalar-vector interactions with higher-$k$ abelian vector shift symmetries.

\vspace{.15cm}
\noindent
{\bf Conventions:}
We work in $D=4$ spacetime dimensions and use the mostly plus metric signature convention.  We work in AdS space with radius $L$, so that $R=-{12/ L^2}<0$, or dS space with Hubble scale $H$.  Expressions can be translated between the two cases with the relation $H^2\leftrightarrow {-1/L^2}$.  
All of our results can be straightforwardly extended to Euclidean signature and to arbitrary dimensions.  Tensors are symmetrized and antisymmetrized with unit weight, e.g., $T_{(\mu\nu)}=\half \left(T_{\mu\nu}+T_{\nu\mu}\right)$ and  $T_{[\mu\nu]}=\half \left(T_{\mu\nu}-T_{\nu\mu}\right)$. We also define the scales $\Lambda_k \equiv \left(m^{k-1} M_{\rm Pl} \right)^{1/k}$.

\section{Symmetries of the free (A)dS spin-1 theory}
\label{sec:freetheory}

Consider a free massive spin-1 field, $A_\mu$, on (A)dS$_4$ with mass $m$, as described by the Proca Lagrangian
\be {\cal L}=  \sqrt{-\bm}\left[ -{1\over 4} F_{\mu\nu}^2 -\frac{m^2}{2} A^2\right]\,, \label{freelage}
\ee
where $F_{\mu\nu}\equiv\partial_\mu A_\nu-\partial_\nu A_\nu$ is the standard Maxwell field strength and indices are raised and lowered with the background (A)dS$_4$ metric, $\bm_{\mu\nu}$.

It was shown in Ref.~\cite{Bonifacio:2018zex} that for particular values of the mass the free theory \eqref{freelage} acquires various shift symmetries.  In particular, at the mass value
\be m^2={6\over L^2}\,,  \label{vecmassvaluee}\ee
the field acquires a symmetry under a shift by a Killing vector $\xi^\mu$,
\be \delta^{(\rm s)} A_{\mu}=\xi_\mu,\quad {\rm where} \quad  \nabla_\mu \xi_\nu+\nabla_\nu \xi_\mu=0\,. \label{labelkillingsyme} \ee
Following the terminology of Ref.~\cite{Bonifacio:2018zex}, we refer to this as the $k=0$ symmetry.

Note that this shift symmetry is nonlinearly realized and is distinct from the unbroken linearly realized (A)dS isometries possessed by any covariant field in an (A)dS background.  The isometries act instead via a Lie derivative,
\be \delta^{(\rm i)}  A_{\mu}=-{\cal L}_\epsilon A_{\mu}=-\left(\epsilon^\nu\nabla_\nu A_\mu+\nabla_\mu \epsilon^\nu\, A_{\nu}\right)\, , \quad \nabla_\mu \epsilon_\nu+\nabla_\nu \epsilon_\mu=0 \,,\label{isometryAfirstdefe}\ee
where $\epsilon^\mu$ is an (A)dS Killing vector, distinct from $\xi^\mu$, parametrizing the (A)dS isometry.

We can understand the presence of the shift symmetry \eqref{labelkillingsyme} by considering the vector as arising from the longitudinal mode of a massive spin-2 field in the massless limit.  The Lagrangian for a spin-2 field, $h_{\mu\nu}$, of mass $m$ in (A)dS space is given by the Fierz--Pauli Lagrangian \cite{Fierz:1939ix} extended to maximally symmetric curved space \cite{Fang:1978se},
\begin{align}
 {\cal L}_m=  \sqrt{-\bm} \Bigg[- \frac{1}{2} \nabla_{\alpha} h_{\mu \nu} \nabla^{\alpha} h^{\mu \nu} &+\nabla_{\alpha} h_{\mu \nu} \nabla^{\nu} h^{\mu \alpha} - \nabla_{\mu} h \nabla_{\nu} h^{\mu \nu} + \frac{1}{2} \nabla_{\mu} h \nabla^{\mu} h  \nn\\
&-{3\over L^2}\left(h_{\mu \nu}h^{\mu \nu} -\frac{1}{2} h^2\right) -\frac{m^2}{2} \left(h_{\mu \nu}h^{\mu \nu} - h^2\right)\Bigg]\, . \label{linearlaggle}
\end{align}
To take the massless limit we introduce a \stu vector $A_\mu$ patterned after the linear diffeomorphism symmetry that is restored in the massless limit,
\be h_{\mu\nu}\mapsto h_{\mu\nu}+\frac{1}{m}\(\nabla_\mu A_\nu+\nabla_\nu A_\mu\)\,, \label{massiveveddcle}
\ee
where the mass scaling is chosen so that the kinetic term for the vector is canonically normalized up to a constant factor.  After the replacement \eqref{massiveveddcle}, the theory has a gauge symmetry
\be
\delta h_{\mu\nu}=\nabla_\mu\xi_\nu+\nabla_\nu\xi_\mu, ~~~~~~~~~~ \delta A_{\mu}=-m\xi_\mu\,, \label{linermadssymde}
\ee
where $\xi_\mu(x)$ is a vector gauge parameter.
We can then take the massless (A)dS limit
\be 
m\rightarrow 0,\quad L\quad {\rm fixed}, \quad {\rm and} \quad h\mn, A_\mu,\xi_\mu \quad {\rm fixed}\,, \label{mmlinelime}
\ee
after which the Lagrangian \eqref{linearlaggle} becomes 
\be
 {\cal L} = {\cal L}_{m=0}+ \sqrt{-\gamma}\left[ -{1\over 2} F_{\mu\nu}^2 - {6\over L^2} A^2\right]\, ,
\ee
where ${\cal L}_{m=0}$ is the Lagrangian~\eqref{linearlaggle} evaluated at $m=0$.
We see the appearance of a massive vector with the mass \eqref{vecmassvaluee}, which carries the three extra degrees of freedom of a massive graviton compared to a massless graviton.

In the massless limit \eqref{mmlinelime}, the gauge symmetry \eqref{linermadssymde} becomes
\be
\delta h_{\mu\nu}=\nabla_\mu\xi_\nu+\nabla_\nu\xi_\mu,~~~~~~~~~~ \delta A_{\mu}=0\,. \label{linermadssymde2}
\ee
This is the transformation that survives the decoupling limit for arbitrary $\xi_\mu$, so the vector field has no gauge symmetry in the limit \eqref{mmlinelime}.  There are, however, special choices of $\xi_\mu$ such that $ A_{\mu}$ does transform after taking the limit.  To look for these, we define
\be 
\hat \xi_\mu=m\xi_\mu, 
\ee
and consider taking the decoupling limit holding $\hat\xi_{\mu}$ fixed,
so that \eqref{linermadssymde} becomes
\be
\delta h_{\mu\nu}={1\over m}\left(\nabla_\mu\hat\xi_\nu+\nabla_\nu\hat\xi_\mu\right),~~~~~~~~~~ \delta A_{\mu}=\hat \xi_\mu\,. \label{linermadssymde3}
\ee
In general, the expression for $\delta h_{\mu\nu}$ in \eqref{linermadssymde3} blows up in the massless limit. However, in the special case where $\hat\xi_\mu$ is a Killing vector, satisfying the Killing equation $\nabla_\mu\hat\xi_\nu+\nabla_\nu\hat\xi_\mu=0,$
then $\delta h_{\mu\nu}$ vanishes and we are left with a finite shift symmetry for $A_\mu$, recovering \eqref{labelkillingsyme}.

\section{Symmetry algebra}
\label{sec:algebra}

In this section we use the ambient space formalism to describe the algebra  formed by the shift symmetries along with the (A)dS isometries and to classify possible deformations.  This gives us information about the possible invariant interactions for shift-symmetric spin-1 theories.

\subsection{Ambient space}
To classify possible symmetries, it is convenient to use the (A)dS ambient space formalism \cite{Dirac:1936fq, Fronsdal:1978vb}, where we describe (A)dS$_4$ as a surface $X^A(x)$ embedded in a $5$-dimensional flat ambient space with Cartesian coordinates $X^A$ (see Appendix B of Ref.~\cite{Bonifacio:2018zex} for more details of the ambient space formalism in our conventions).  A vector field $A_\mu(x)$ with mass~\eqref{vecmassvaluee} is bijectively mapped to a vector field, $A_B(X)$, in the ambient space satisfying the following homogeneity and transversality conditions:
\be  
X^A\partial_A A_B= A_B\quad{\rm and}\quad  X^BA_B=0,\label{Aabmientrelationse}
\ee
where the (A)dS vector is recovered via the pullback,
\be
A_\mu(x)=\partial_\mu X^B(x) A_B  \left( X(x) \right).
\ee

(A)dS Killing vectors are represented in ambient space by constant antisymmetric tensors.  We call the antisymmetric tensor associated to the (A)dS isometries $J_{AB}$. The Killing vector $\epsilon^\mu$ of \eqref{isometryAfirstdefe} is then given through the relation 
\be
\epsilon_\mu(x)=\partial_\mu X^A(x)J_{AB}X^B(x). 
\ee
We call the antisymmetric tensor associated to the shift symmetries $S_{AB}$, so the Killing vector $\xi^\mu$ of \eqref{labelkillingsyme} is given through the analogous relation
\be
\xi_\mu(x)=\partial_\mu X^A(x)S_{AB}X^B(x).
\ee

\subsection{Abelian shift symmetry\label{abeliansubsecs}}

We now describe the symmetry algebra of the free theory.
The isometries acting on the ambient space vector take the form
\be
\delta_{J_{AB}}A_{C}\equiv J_{AB}A_{C}=\left(X_A\partial_B-X_B\partial_A\right)A_{C}+\left({\cal J}_{AB}\right)_{C}^{\ D}A_{D},
\label{eq:adssymmtrans}
\ee
where $\left({\cal J}_{AB}\right)_{C}{}^{D}\equiv\eta_{AC}\delta_B{}^{D}-\eta_{BC}\delta_A{}^{D}$ is the Lorentz generator in the vector representation. 
These satisfy the commutation relations of the (A)dS$_4$ isometry algebra, $\mathfrak{so}(5)$,
\be \left[ J_{AB},J_{CD}\right]= \eta_{AC}J_{BD}-\eta_{BC}J_{AD}+\eta_{BD}J_{AC}-\eta_{AD}J_{BC}.\label{adsalgebrae}\ee
The shift symmetry \eqref{labelkillingsyme} written in ambient space is $\delta A_B=S_{BC}X^C$, so its form is
\be
\delta_{S_{AB}}A_{C}\equiv S_{AB}A_{C}=\eta_{C[A }X_{B]}   .
\label{eq:shiftsymmtrans}
\ee
Due to the (A)dS covariance of the shift generators $S_{AB}$, their commutators with the isometries \eqref{eq:adssymmtrans} are 
\be \left[ J_{AB},S_{CD}\right]= \eta_{AC}S_{BD}-\eta_{BC}S_{AD}+\eta_{BD}S_{AC}-\eta_{AD}S_{BC}.\label{ssalgebrae}\ee
Since the shifts \eqref{eq:shiftsymmtrans} are independent of the field, they commute among themselves,
\be \left[ S_{AB},S_{CD}\right]= 0.\label{sscommutatorees}\ee
The commutators \eqref{adsalgebrae}, \eqref{ssalgebrae}, and \eqref{sscommutatorees} together close to form the algebra of the free theory.  It is a semi-direct product of $\mathfrak{so}(5)$ with the abelian algebra of the shifts.

\subsection{Non-abelian deformations\label{nonabeliansec}}

We now look for deformations of the free symmetry algebra. The action of the isometries on the field is unchanged in an interacting theory, so the commutators \eqref{adsalgebrae} remain the same.  

The shift symmetry of the vector may be deformed by terms involving powers of the fields, 
\be
\delta_{S_{AB}}A_{C}\equiv S_{AB}A_{C}=\eta_{C[A }X_{B]}+{\cal O}\left(A\right) .
\label{eq:shiftsymmtrans2}
\ee 
Regardless of the deformation, the commutator \eqref{ssalgebrae} remains the same since the symmetry \eqref{eq:shiftsymmtrans2} should remain (A)dS covariant. However, the commutator \eqref{sscommutatorees} can be modified.  The most general structure which can appear on the right-hand side that is consistent with the symmetries of the left-hand side is
\begin{align} \left[S_{AB}, S_{CD}\right] &=  {\alpha} \left( \eta_{AC}J_{BD}-\eta_{BC}J_{AD}+\eta_{BD}J_{AC}-\eta_{AD}J_{BC}\right) \nn\\
&+{\beta} \left( \eta_{AC}S_{BD}-\eta_{BC}S_{AD}+\eta_{BD}S_{AC}-\eta_{AD}S_{BC}\right).\label{scommutatore2}
\end{align}
This ansatz satisfies all Jacobi identities for any values of the constants $\alpha$ and $\beta$, so it describes a consistent algebra.

When $\beta^2+4\alpha=0$, this reduces to the abelian algebra of the free theory, as can be seen by defining
\be
\tilde{S}_{AB} \equiv S_{AB} - \frac{\beta}{2} J_{AB},
\ee
which commutes with itself.
For any other values of $\alpha$ and $\beta$, we can make the following change of basis:
\be 
B^{\pm}_{AB}=\frac{\sqrt{4 \alpha +\beta ^2}\pm\beta }{2 \sqrt{4 \alpha +\beta ^2}}J_{AB}  \mp \frac{1}{\sqrt{4 \alpha +\beta ^2}} S_{AB},
\ee
which has inverse
\be 
J_{AB}=B^{+}_{AB}+B^{-}_{AB},~~~~~~~~ S_{AB}={\beta -\sqrt{4 \alpha +\beta ^2}\over 2}B^{+}_{AB}+{\beta +\sqrt{4 \alpha +\beta ^2}\over 2}B^{-}_{AB},
\ee
after which we see that the algebra is $\mathfrak{so}(5)\oplus \mathfrak{so}(5)$,
\begin{align} & \left[ B^+_{AB},B^+_{CD}\right]= \eta_{AC}B^+_{BD}-\eta_{BC}B^+_{AD}+\eta_{BD}B^+_{AC}-\eta_{AD}B^+_{BC},\nn\\ 
& \left[ B^-_{AB},B^-_{CD}\right]= \eta_{AC}B^-_{BD}-\eta_{BC}B^-_{AD}+\eta_{BD}B^-_{AC}-\eta_{AD}B^-_{BC},\nn\\ 
& \left[ B^+_{AB},B^-_{CD}\right]= 0.
\end{align}
Since the $J_{AB}$ are unbroken symmetries and the $S_{AB}$ are broken symmetries, the symmetry breaking pattern breaks two copies of the (A)dS group down to the  diagonal subgroup,
\be \mathfrak{so}(5)\oplus \mathfrak{so}(5)\rightarrow \mathfrak{so}(5)_{\rm diag}.\label{breakingpatternse}\ee

A specific deformation of the form \eqref{eq:shiftsymmtrans2} that realizes the non-abelian algebra is\footnote{One way to find this is to look for modifications to the shift symmetry that involve at most one derivative.  The only possible terms which preserve the homogeneity condition $X^A\partial_A A_B= A_B$ and close with the isometries under commutation are then given by 
\be
S_{AB}A_C=\eta_{C[A}X_{B]}+\beta_1 \eta_{C[A}A_{B]}+\beta_2 X_{[A}\partial_{B]}A_C+4\alpha A_{[A}\partial_{B]}A_C\, ,
\ee
with free coefficients $\beta_1$, $\beta_2$, and $\alpha$.  (Note there are no terms with higher powers of $A$ consistent with the scaling requirement without introducing more derivatives.)  Requiring that this preserves the transversality condition $X^BA_B=0$ implies that $\beta_1=\beta_2$.  The terms proportional to the $\beta$'s are then nothing but an (A)dS isometry \eqref{eq:adssymmtrans}, so we can drop them without loss of generality, which leads to \eqref{canparanonlinee}.}
\be S_{AB}A_C=\eta_{C[A}X_{B]}+4\alpha A_{[A}\partial_{B]}A_C\,, \label{canparanonlinee}\ee
which gives the commutator \eqref{scommutatore2} with $\beta=0$. We could now proceed to directly search for (A)dS interactions that are invariant under the restriction of the transformation \eqref{canparanonlinee} to the (A)dS surface. However, in what follows it will be more convenient to construct the interacting theory from massive gravity, which leads to a different parametrization of the symmetry transformation.

\section{Spin-1 theory with abelian shift symmetry from pseudo-linear interactions}
\label{sec:abelian}

A simple way to construct interacting spin-1 theories realizing the abelian algebra of Section \ref{abeliansubsecs} is to consider pseudo-linear interactions of a massive spin-2 particle.  These are constructed by starting with the linear Lagrangian \eqref{linearlaggle} and adding nonlinear potential terms,
\be {\cal L}={\cal L}_m+ \sqrt{-\bm}m^2 M_{\rm Pl}^2 V(h/M_{\rm Pl}) \,,\label{plmggffe}
\ee
where indices are contracted using the background (A)dS metric, $\bm_{\mu\nu}$. Here $M_{\rm Pl}$ is an interaction scale analogous to the Planck mass and $V(h/M_{\rm Pl})$ is an analytic function of $h_{\mu\nu}/M_{\rm Pl}$ that starts at cubic order in the field (the mass terms are already included in ${\cal L}_m$). 

We now introduce a \stu field through the same linear replacement \eqref{massiveveddcle} as in the linear theory, which leaves the massless kinetic term invariant.  This leaves the theory invariant under the linearized gauge symmetry \eqref{linermadssymde}.  We may then take a massless decoupling limit, keeping fixed the (A)dS scale and the strong coupling scale $\Lambda_2\equiv (M_{\rm Pl} m)^{1/2}$,
\be 
m\rightarrow 0,\quad M_{\rm Pl}\rightarrow \infty,~~~{\rm with}~~~ L, \, \Lambda_2~~{\rm and}~~ h\mn,\, A_\mu \quad {\rm held~fixed}\,. \label{mmlinelime2}
\ee
At nonlinear orders, this decoupling limit amounts to keeping only the $A_\mu$ terms from the potential.  Thus the decoupling limit consists of a decoupled free massless spin-2 field and a self-interacting vector Lagrangian given by
\be 
{\cal L}_A= \sqrt{-\gamma}\left[ -{1\over 2} F_{\mu\nu}^2 - {6\over L^2} A_{\mu}^2+\Lambda_2^4 V(B/\Lambda_2) \right]\,, \label{plvtvacde}
\ee
where
\be B_{\mu\nu}\equiv  \nabla_\mu A_\nu+\nabla_\nu A_\mu\,. \label{massiveveddcle2}
\ee
The Lagrangian is a nonlinear function only of the tensor $B_{\mu \nu}$ which is manifestly invariant under the shift symmetry of the linear theory, \eqref{labelkillingsyme}.  The linear part of the Lagrangian, which comes from the Fierz--Pauli mass term, can also be written in terms of $B_{\mu\nu}$ as $\sim B_{\mu\nu}^2-B^2$. 

The theory \eqref{plvtvacde} is generally ghostly, but we can write a ghost-free version by using the pseudo-linear potential extended to (A)dS space \cite{Folkerts:2011ev,Hinterbichler:2013eza, Akagi:2014iea},
\be V(h) =-\frac{\alpha_3}{2}\epsilon^{\mu_1\mu_2\mu_3\lambda} \epsilon^{\nu_1\nu_2\nu_3}_{\ \ \ \ \  \ \ \lambda} h_{\mu_1\nu_1} h_{\mu_2\nu_2} h_{\mu_3\nu_3} -\frac{\alpha_4}{2}\epsilon^{\mu_1\mu_2\mu_3\mu_4} \epsilon^{\nu_1\nu_2\nu_3\nu_4}_{} h_{\mu_1\nu_1} h_{\mu_2\nu_2} h_{\mu_3\nu_3} h_{\mu_4\nu_4} \,,\label{pseudolinearaddse}
\ee
where $\alpha_3$, $\alpha_4$ are two free parameters. Replacing $h_{\mu \nu} \mapsto B_{\mu \nu}$ gives a ghost-free vector theory.

The above construction tells us that $B_{\mu\nu}$ is an invariant building block out of which we can make shift-symmetric interacting theories.  In these theories the symmetry algebra is undeformed from the linear case, so in this sense they are analogous to the galileons or (A)dS galileons \cite{Goon:2011qf,Goon:2011uw,Burrage:2011bt}. 

Now that we have the invariant building block, we can remove the scaffolding of the massive spin-2 construction and consider more general invariant Lagrangians given by any scalar function $F$ of $B_{\mu\nu}$ and the background covariant derivative $\nabla_\mu$,
\be
{\cal L}_A= \sqrt{-\bm}F(B,\nabla)\,,
\ee
and the result will be an interacting spin-1 Lagrangian that is trivially invariant under \eqref{labelkillingsyme}.

\section{Spin-1 theory with non-abelian shift symmetry from massive gravity\label{nonlinemgsec}}
With the experience gained from studying the interacting theory that realizes the abelian algebra, 
we can now guess that theories invariant under the non-abelian algebra of Section \ref{nonabeliansec} will be found by starting with nonlinear massive gravity built from the Einstein--Hilbert term,  
\be
{\cal L}=  {M_{\rm Pl}^2\over 2} \sqrt{-g}\left[R(g)+{6\over L^2}+m^2 V(g,\bm)\right].\label{genghostlagmge}
\ee
Here $V$ is a scalar function of $g_{\mu\nu}$, which is the full dynamical metric, and $\bm_{\mu\nu}$, which is the background (A)dS metric.

The nonlinear \stu procedure on (A)dS space can be carried out by replacing \cite{deRham:2012kf}\footnote{See Appendix C of Ref.~\cite{deRham:2018svs} for a summary in our conventions. A covariant \stu formalism was also studied in Refs.~\cite{Gao:2014ula,Gao:2015xwa}. }
\be
\bm_{\mu\nu}\mapsto \tilde{\gamma}_{\mu \nu}={\gamma}_{\mu \nu}-S_{\mu\nu}-S_{\nu\mu}+S_{\mu}^{\ \lambda}S_{\nu\lambda}-{1\over L^2+A^2}T_{\mu} T_{\nu}\,, \label{stuckreplacee}
\ee
where
\be
S_{\mu\nu}\equiv   \nabla_\mu A_\nu+\gamma_{\mu\nu}\left(1-\sqrt{1+{1\over L^2}A^2}\right),\quad  T_{\mu}\equiv {1\over 2}\partial_\mu (A^2)-\sqrt{1+{1\over L^2}A^2}A_\mu\,, \label{stuckreplacee2}
\ee
and where indices in these expressions are raised and lowered using $\bm_{\mu\nu}$. 

Next we expand in metric perturbations,
\be
g_{\mu\nu}=\gamma_{\mu\nu}+h_{\mu\nu},
\ee
and define the fields
\be
\hat h_{\mu\nu}= {M_{\rm Pl}\over 2 } h_{\mu\nu},  \quad  \hat A_\mu= {M_{\rm Pl} m\over 2} A_\mu  \,, \label{cannormee1}
\ee
which are canonically normalized up to a constant factor.
We then take the following decoupling limit with $\Lambda_2$ fixed,
\be
 M_{\rm Pl} \rightarrow \infty,\quad m\rightarrow 0,~~~{\rm with}~~~ L,\, \Lambda_2~~ {\rm and}~~ \hat h\mn,\,  \hat A_\mu \quad  {\rm held~fixed.}\label{declimitge}
\ee
In this decoupling limit, all $\hat{h}_{\mu\nu}$'s coming from the potential, as well as nonlinear terms in the Einstein--Hilbert action, scale to zero, leaving a decoupled linear massless spin-2 field and the following self-interacting vector theory:
\be
{\cal L}_A={ \Lambda_2^4 \over 2}\sqrt{-\bm}V(\bm,\tilde \bm)\,, \label{nonlinegernthere}
\ee
with $\tilde\bm_{\mu\nu}$ as in \eqref{stuckreplacee}.

The theory \eqref{nonlinegernthere} will generally have ghosts but---as discussed in Ref.~\cite{deRham:2018svs}---it will be ghost free if we choose the potential to be that of dRGT massive gravity \cite{deRham:2010kj}  (see Refs.~\cite{Hinterbichler:2011tt,deRham:2014zqa} for reviews).  On an (A)dS background \cite{Hassan:2011tf},  the dRGT Lagrangian is
\be 
{\cal L}={M_{\rm Pl} ^2\over 2}\sqrt{-g}\left( R[g]+{6\over L^2} +{m^2}\left[S_2({\cal K})+\alpha_3 S_3({\cal K})+\alpha_4 S_4({\cal K})\right]\right)\,, \label{nonlinlag}
\ee
where the tensor ${\cal K}^\mu_{\ \nu}$ is defined by
\be {\cal K}^\mu_{\ \nu}=\delta^\mu_{\ \nu}-\left(\sqrt{g^{-1} \gamma}\right)^\mu_{\ \nu}\, ,\ee
and $S_n$ are the symmetric polynomials, defined by $S_n^{\rm }({\cal K})=n!\, {\cal K}^{[\mu_1}_{\ \mu_1}{\cal K}^{\mu_2}_{\ \mu_2}\cdots {\cal K}^{\mu_n]}_{\ \mu_n}$.
As for the ghost-free pseudo-linear potentials, there are two dimensionless free parameters, $\alpha_3$ and $\alpha_4$.

\subsection{Shift symmetry of the decoupling limit vector theory}

We now show that the interacting spin-1 theory~\eqref{nonlinegernthere}, which appears in the decoupling limit of massive gravity, is automatically invariant under a nonlinear shift symmetry that realizes the algebra of Section~\ref{nonabeliansec}.

After the \stu fields are introduced, but before taking any decoupling limit, the massive gravity action has the diffeomorphism symmetry that the \stu fields are designed to introduce.  This symmetry reads (see
appendix~\ref{projapp} for the derivation) 
 \begin{align}
\delta h_{\mu\nu}&=\nabla_\mu\xi_\nu+\nabla_\nu\xi_\mu+\mathcal{L}_{\xi}h_{\mu\nu}\, ,\\
\delta A_{\mu} &=-\xi_{\mu}+\xi^{\rho}\nabla_{\rho}A_{\mu}+\xi_{\mu}\left(1-\sqrt{1+{1\over L^2}A^2}\right)\, . \label{diffbeforenorme}
\end{align}
In terms of the hatted fields defined in Eq.~\eqref{cannormee1},  
this becomes
\begin{align}
\delta \hat h_{\mu\nu}&=\frac{M_{\rm Pl}}{2}\left(\nabla_\mu\xi_\nu+\nabla_\nu\xi_\mu\right)+\mathcal{L}_{\xi}\hat h_{\mu\nu}\, ,\\
\delta \hat A_{\mu}&=-\frac{mM_{\rm Pl}}{2}\xi_{\mu}+\xi^{\rho}\nabla_{\rho}\hat A_{\mu}+\frac{mM_{\rm Pl}}{2}\xi_{\mu}\left(1-\sqrt{1+{4 \hat A^2\over (mM_{\rm Pl} L)^2}}\right). \label{eqshiftermsse}
\end{align}

To see the gauge symmetry of the decoupling limit, we normalize $\xi_{\mu}={2  \over M_{\rm Pl}}\hat \xi_{\mu}$ and then take the limit \eqref{declimitge} with $\hat \xi_{\mu}$ fixed. This leaves only 
\be \delta \hat h_{\mu\nu}=\nabla_\mu\hat\xi_\nu+\nabla_\nu\hat\xi_\mu,\ \ \ \delta \hat A_\mu=0,\ee
which is the linear diffeomorphism symmetry of the decoupled linear kinetic term for $h_{\mu\nu}$.

The shift symmetry we seek, on the other hand, has as its leading part the first term in \eqref{eqshiftermsse}. To see this global symmetry we define $\bar{\xi}_{\mu}=\frac{1}{2}mM_{\rm Pl}\xi_{\mu}$, so the transformations become
\begin{align}
\delta\hat{h}_{\mu\nu}&={1\over m}\left(\nabla_\mu\bar{\xi}_\nu+\nabla_\nu\bar{\xi}_\mu\right)+\frac{2}{m M_{\rm Pl}}\mathcal{L}_{\bar{\xi}}\hat{h}_{\mu\nu} \, ,\label{firsttermsere}\\
\delta \hat{A}_{\mu}&=-\bar{\xi}_{\mu}+\frac{2}{mM_{\rm Pl}}\bar{\xi}^{\rho}\nabla_{\rho}\hat{A}_{\mu}+\bar{\xi}_{\mu}\left(1-\sqrt{1+{4\hat{A}^2\over {(mM_{\rm Pl} L)}^2}}\right).
\end{align}
Now in the decoupling limit \eqref{declimitge} with $\bar \xi_{\mu}$ fixed, the first terms in \eqref{firsttermsere} blow up unless $\bar\xi_\mu$ is a Killing vector, for which $\nabla_\mu\bar{\xi}_\nu+\nabla_\nu\bar{\xi}_\mu=0$.  In this case we get a finite limit which is the nonlinear global symmetry,
\begin{align}
\delta\hat{h}_{\mu\nu}&=\frac{2}{\Lambda_2^2}\mathcal{L}_{\bar{\xi}}\hat{h}_{\mu\nu} \,, \label{firsttermsere2}\\
\delta \hat{A}_{\mu}&=-\bar{\xi}_{\mu}+\frac{2}{\Lambda_2^2}\bar{\xi}^{\rho}\nabla_{\rho}\hat{A}_{\mu}+\bar{\xi}_{\mu}\left(1-\sqrt{1+{4\hat{A}^2\over {(\Lambda_2^2 L)}^2}}\right) \nn\\
&=-\bar{\xi}_{\mu}-\frac{2}{\Lambda_2^2}\nabla_{\mu}\bar{\xi}^{\nu}\hat{A}_{\nu}+\bar{\xi}_{\mu}\left(1-\sqrt{1+{4\hat{A}^2\over {(\Lambda_2^2 L)}^2}}\right)+\frac{2}{\Lambda_2^2}\mathcal{L}_{\bar{\xi}}\hat{A}_{\mu}\,.
 \label{procashiftinte}
\end{align}
The Lie derivative terms in $\delta \hat{h}_{\mu\nu}$ and $\delta \hat{A}_{\mu}$ are simply an (A)dS isometry, so we can remove them by redefining the symmetry transformations to subtract off these (A)dS isometries.
 This leaves a non-linear shift symmetry that acts only on the vector field, which simplifies to
 \be \delta \hat{A}_{\mu}=-\frac{2}{\Lambda_2^2}\nabla_{\mu}\bar{\xi}^{\nu}\hat{A}_{\nu}-\bar{\xi}_{\mu}\sqrt{1+{4\hat{A}^2\over {(\Lambda_2^2 L)}^2}} \,.\label{nowvectsymmk2}
\ee
We see explicitly how the global part of the diffeomorphism symmetry gets inherited by the vector field as a shift symmetry. 

Upon going back to non-canonical normalization, the shift symmetry \eqref{nowvectsymmk2} reads
\begin{equation}
\delta^{(\rm s)}_{\xi}A_{\mu}=-\nabla_{\mu}{\xi}^{\nu}{A}_{\nu}-\xi_{\mu}\sqrt{1+\frac{A^2}{L^2}}.\label{finalshiftsym}
\end{equation}
Under this transformation, the \stu metric $\tilde\gamma_{\mu\nu}(A)$ defined in \eqref{stuckreplacee} transforms as a tensor,
\be 
\delta^{(\rm s)}_{\xi}\tilde\gamma_{\mu\nu}=  \mathcal{L}_{{\xi}} \tilde\gamma_{\mu\nu}. 
\ee 
We can therefore use it, disregarding its origin from the decoupling limit of massive gravity, as a covariant building block to make invariant vector theories.  Any covariant Lagrangian made from $\tilde\gamma_{\mu\nu}$ and the background (A)dS metric $\gamma_{\mu\nu}$ will be invariant under the shift symmetry \eqref{finalshiftsym},
\be {\cal L}=\sqrt{-\gamma}F(\gamma,\tilde\gamma,\nabla,\tilde\nabla,\cdots).\ee
This works because $\xi_\mu$ is an (A)dS Killing vector, so that we have $\mathcal{L}_{{\xi}} \gamma_{\mu\nu}=0$.

\subsection{Algebra of symmetries}
\noindent

We now proceed to calculate the commutation relations among the nonlinear shift symmetries \eqref{finalshiftsym} and the background (A)dS isometries \eqref{isometryAfirstdefe}, verifying that they satisfy the unique deformed algebra identified in Section \ref{nonabeliansec}.
Directly computing the commutators gives the following:
\begin{itemize}
\item
The commutator of two isometries is
\be
\big[\delta^{(\rm i)}_{\epsilon},\delta^{(\rm i)}_{\epsilon '}\big]A_{\mu}=\delta^{(\rm i)}_{[\epsilon,\epsilon ']}A_{\mu}\, ,
\ee
where $[\epsilon,\epsilon ']^\mu={\epsilon }^{\nu}\nabla_{\nu}{\epsilon'}^{\mu}-{\epsilon' }^{\nu}\nabla_{\nu}\epsilon^{\mu}$ is the Lie bracket. 
This is the $\mathfrak{so}(5)$ algebra of (A)dS isometries.
 
\item

The commutator of a shift symmetry with an isometry is another shift symmetry,
\begin{align}
\big[\delta^{(\rm i)}_{\epsilon},\delta^{(\rm s)}_{\xi}\big]A_{\mu}&=\delta^{(\rm s)}_{[\epsilon,\xi]}A_{\mu}\,.
\end{align}
This tells us that the shifts transform covariantly under isometries.

\item

The commutator of two shift symmetries is again a shift symmetry
\begin{align}
\big[\delta^{(\rm s)}_{\xi},\delta^{(\rm s)}_{\xi '}\big]A_{\mu}&=\delta^{(\rm s)}_{[\xi,\xi ']}A_{\mu}\,.
\end{align}

\end{itemize}
To obtain all these commutators we must use the fact that the $\epsilon$'s and $\xi$'s satisfy Killing's equation.

If we now define the linear combinations 
\begin{align}
\delta^{(+)}_{\xi}A_{\mu}&\equiv\left(\delta^{({\rm i})}_{\xi}-\delta^{({\rm s})}_{\xi}\right)A_\mu 
=-\xi^{\nu}\nabla_{\nu}A_{\mu}+\xi_{\mu}\sqrt{1+\frac{A^2}{L^2}}\, ,\\ 
\delta^{(-)}_{\xi}A_{\mu}&\equiv \delta^{({\rm s})}_{\xi}A_\mu 
=-\nabla_{\mu}\xi^{\nu}A_{\nu}-\xi_{\mu}\sqrt{1+\frac{A^2}{L^2}}\, ,
\end{align}
then $\delta^{(\pm)}_{}$ satisfy
\begin{align}
\big[\delta^{(+)}_{\xi},\delta^{(+)}_{\xi'}\big]A_{\mu}&=\delta^{(+)}_{[\xi,\xi ']}A_{\mu}\, ,\\
\big[\delta^{(-)}_{\xi},\delta^{(-)}_{\xi'}\big]A_{\mu}&=\delta^{(-)}_{[\xi,\xi ']}A_{\mu}\, ,\\
\big[\delta^{(+)}_{\xi},\delta^{(-)}_{\xi'}\big]A_{\mu}&=0\, ,
\end{align}
which are the commutators of $\mathfrak{so}(5)\oplus \mathfrak{so}(5)$ with the breaking pattern \eqref{breakingpatternse}, so that the diagonal $\mathfrak{so}(5)$ is linearly-realized.

\section{Flat space symmetries}
\label{sec:flatspace}

We now study the flat space limits of the various interacting (A)dS vector theories that we have been considering.  In the flat limit, the mass \eqref{vecmassvaluee} goes to zero and the massive spin-1 particle decomposes into a massless spin-1 particle plus a massless scalar that describes the longitudinal mode. 

The resulting theories can also be obtained as the scalar-vector sector of the flat space decoupling limits of massive gravity or the massive spin-2 pseudo-linear theory.  In the case of dRGT massive gravity \eqref{nonlinlag}, the flat space decoupling limit is
\be
m\rightarrow 0, \quad  L\rightarrow \infty,\quad M_{\rm Pl}\rightarrow \infty, \quad {\rm with}\quad m L, \,\Lambda_3 \quad {\rm held~fixed},
\ee
where $\Lambda_3 \equiv (M_{\rm Pl}m^2)^{1/3}$. The decoupling limit contains three types of interactions: scalar self-interactions in the form of galileons; scalar-tensor interactions with one power of $h_{\mu\nu}$ and various powers of $\partial_\mu\partial_\nu\phi$, whose coefficients all depend on $mL$ \cite{deRham:2012kf}; and scalar-vector interactions with two powers of $\partial_\mu A_\nu$ and various powers of $\partial_\mu\partial_\nu\phi$ that do not depend on $mL$. The only important part for us is the scalar-vector interactions, which have the following schematic structure:\footnote{The full decoupling limit including the vectors was derived in Refs.~\cite{Gabadadze:2013ria,Ondo:2013wka}, while partial results can be found in Refs.~\cite{deRham:2010gu,Koyama:2011wx,Tasinato:2012ze,Yu:2013owa}. An explicit expression up to quartic order in the fields is also given in Eq.~(4.20)  of Ref.~\cite{deRham:2018svs}.}
\begin{align} {\cal L}_{ \rm mg}(\hat{A}, \hat{\phi}) = &-6 (\partial \hat{\phi})^2-{1\over 2}\hat{F}_{\mu\nu}^2+{\cal O}\left(\hat F^2\left[\hat\Pi+ \hat \Pi^2+\cdots\right]\right) \nn\\
&+
{\left(3 \alpha_3 +1 \right) \over 2 \Lambda_3^3}\epsilon^{\mu_1\mu_2\mu_3\lambda} \epsilon^{\nu_1\nu_2\nu_3}{}{}{}_{\lambda}\hat  F_{\mu_1\mu_2} \hat F_{\nu_1\nu_2} \hat \Pi_{\mu_3\nu_3} +{\cal O}\left(\hat F^2\left[\hat\Pi^2+ \hat \Pi^3+\cdots\right]\right)\label{fulldrgtdecouplesve}\\
&+ \frac{\left(12\alpha_4 +3 \alpha_3\right)}{2 \Lambda_3^6}\epsilon^{\mu_1\mu_2\mu_3\mu_4} \epsilon^{\nu_1\nu_2\nu_3\nu_4}_{} \hat F_{\mu_1\mu_2} \hat F_{\nu_1\nu_2} \hat \Pi_{\mu_3 \nu_3}\, \hat \Pi_{\mu_4 \nu_4} +{\cal O}\left(\hat F^2\left[\hat \Pi^3+\cdots\right]\right)\, ,\nn 
\end{align}
where $\hat{F}_{\mu\nu}\equiv\partial_\mu \hat{A}_\nu-\partial_\nu \hat{A}_\mu$ and $\hat{\Pi}_{\mu \nu} \equiv \partial_{\mu} \partial_{\nu} \hat{\phi}$. The hatted fields, which are canonically normalized up to a constant factor, are rescaled \stu fields that are defined below. The reason for writing the interactions as in Eq.~\eqref{fulldrgtdecouplesve} is also explained below.

{For a generic ghostly theory of massive gravity of the form \eqref{genghostlagmge}, the strong coupling scale is some scale $\Lambda_*<\Lambda_3$. The $\Lambda_*$ decoupling limit consists of a decoupled free tensor and vector plus a scalar with self interactions which are functions of second derivatives of the scalar suppressed by $\Lambda_*$ \cite{Deffayet:2005ys}.}

The decoupling limit of massive gravity is invariant under the galileon shift symmetry \eqref{eq:galshift}, and this is the only global symmetry known so far. However, since the scalar-vector interactions \eqref{fulldrgtdecouplesve} are independent of $mL$, they must also appear in the flat space limit of the (A)dS spin-1 theory with the non-abelian shift symmetry, which is itself a decoupling limit of massive gravity.
 This raises a puzzle: our massive spin-1 theory has 10 broken symmetries, one for each independent (A)dS$_4$ Killing vector, but the flat limit seems to have fewer, four in the galileon symmetry $b_\mu$ and one in the shift symmetry $c$.  What happened to the other symmetries?   
 
 As we will see, the other symmetries are indeed present, in the form of a shift of the helicity-1 mode,
\be \delta \hat{A}_\mu=m_{\mu\nu}\left(x^\nu-{2\over \Lambda_3^3} \partial^\nu\hat \phi\right)\, ,\label{vectsyminpe}\ee
where $m_{\mu\nu}$ is a constant antisymmetric tensor.  
In the flat limit, the 10 broken (A)dS transformations decompose into Poincar\'e transformations. The galileon shifts $b^\mu$ correspond to the translations and $m_{\mu\nu}$ correspond to the Lorentz transformations, which accounts for the missing symmetries.  Including the additional six Lorentz transformations, the flat limit now has 11 symmetries, which is one more than the massive theory.  The extra symmetry is the shift symmetry $c$, which is the generic extra shift symmetry that the longitudinal mode of a vector acquires in its massless limit.

There is a similar story for the (A)dS theory with the abelian shift symmetry, whose flat space limit corresponds to the scalar-vector sector of the decoupling limit of the massive spin-2 pseudo-linear theory and which has a free version of the vector symmetry \eqref{vectsyminpe}, which acts as
\be 
\delta { A}_\mu=m_{\mu\nu}x^\nu.
\ee
We now show explicitly how these symmetries arise from two perspectives: directly in the massive spin-2 decoupling limits, and from the flat-space limits of the (A)dS shift-symmetric vector theories.

\subsection{Massive gravity decoupling limit\label{symfromflatssection}}

We start by deducing the existence of the new symmetry \eqref{vectsyminpe} directly from the standard flat space \stu replacement and decoupling limit procedure as applied to massive gravity. 

The \stu procedure for massive gravity on flat space consists of the following replacement on the metric fluctuation away from the fiducial flat metric, $H_{\mu\nu} \equiv g_{\mu\nu}-\eta_{\mu\nu}$ \cite{ArkaniHamed:2002sp, deRham:2010ik},
\be 
H_{\mu\nu}\mapsto  h_{\mu\nu}+\partial_\mu A_\nu+\partial_\nu A_\mu-\partial_\mu A^\alpha \partial_\nu A_\alpha \,. 
\ee
This decomposition has the \stu gauge invariance
\begin{align}
\delta h_{\mu\nu}&=\partial_\mu \xi_\nu+\partial_\nu \xi_\mu+{\cal L}_\xi h_{\mu\nu}, \\
 \delta A_\mu&=-\xi_\mu+\xi^\nu\partial_\nu A_\mu\,. \label{gaugesyms1ene}
 \end{align}
We then make a further \stu replacement to introduce a scalar
\be A_{\mu}\mapsto A_{\mu}+\partial_\mu\phi,\ee
which gives the gauge symmetry,\footnote{Note that there is a typo in earlier versions of Ref.~\cite{Hinterbichler:2011tt}: the $\phi$-dependent term on the right-hand side of $\delta A_\mu$ was missing, and this is crucial for the new symmetry.}
\begin{align} \delta h_{\mu\nu}&=\partial_\mu \xi_\nu+\partial_\nu \xi_\mu+{\cal L}_\xi h_{\mu\nu}, \\
 \delta A_\mu&=\partial_\mu\Lambda-\xi_\mu+\xi^\nu\partial_\nu (A_\mu+\partial_\mu\phi), \\ 
 \delta \phi&=- \Lambda. \label{gaugesyms2}
 \end{align}
Defining  the fields
\be \label{canonicalfields}
\hat{h}_{\mu \nu}=\half M_{\rm Pl} h_{\mu \nu},\quad \hat A_{\mu}=\half mM_{\rm Pl} A_{\mu}, \quad \hat \phi=\half m^2 M_{\rm Pl}\phi,
\ee
which are canonically normalized up to a constant factor,
the gauge transformations take the form
\begin{align}
\delta \hat h_{\mu\nu}&={1\over 2}{M_{\rm Pl}}\left(\partial_\mu \xi_\nu+\partial_\nu \xi_\mu\right)+{\cal L}_\xi \hat h_{\mu\nu}, \nn\\
 \delta \hat A_\mu&={1\over 2}m M_{\rm Pl}\, \partial_\mu\Lambda-{1\over 2}mM_{\rm Pl}\,\xi_\mu+\xi^\nu\partial_\nu \left(\hat A_\mu+{1\over m} \partial_\mu\hat \phi\right), \nn\\ 
 \delta \hat \phi&=- {1\over 2}m^2M_{\rm Pl}\Lambda\,. \label{gaugesyms2a}
 \end{align}

To investigate the gauge symmetry of the decoupling limit, we normalize $\hat \xi_\mu={1\over 2}M_{\rm Pl}\xi_{\mu}$ and $\hat \Lambda={1\over 2}m M_{\rm Pl}\Lambda$ so the fields transform as
\begin{align}
\delta \hat h_{\mu\nu}&=\partial_\mu \hat \xi_\nu+\partial_\nu\hat \xi_\mu+{2\over M_{\rm Pl}}{\cal L}_{\hat\xi} \hat h_{\mu\nu}, \nn\\
 \delta \hat A_\mu&=\partial_\mu\hat \Lambda-m\,\hat\xi_\mu+{2\over M_{\rm Pl}}\hat\xi^\nu\partial_\nu \left(\hat A_\mu+{1\over m} \partial_\mu\hat \phi\right), \nn\\ 
 \delta \hat \phi&=- m\hat\Lambda. \label{gaugesyms2b}
 \end{align}
 In the decoupling limit with $m\rightarrow 0$, $M_{\rm Pl}\rightarrow \infty$, any scale smaller than $\Lambda_2$ held fixed, and all the hatted fields and gauge parameters held fixed, this reduces to the linear gauge symmetry
\begin{align}
\delta \hat h_{\mu\nu}&=\partial_\mu \hat \xi_\nu+\partial_\nu\hat \xi_\mu\, ,\nn\\
 \delta \hat A_\mu&=\partial_\mu\hat \Lambda , \nn\\ 
 \delta \hat \phi&=0\,.\label{gaugesyms2c}
 \end{align}
 These are the only symmetries that survive the decoupling limit for arbitrary $\Lambda$ and $\xi_\mu$.
 
For specific choices of $\Lambda$ and $\xi_\mu$, corresponding to reducibility parameters of the gauge symmetries, there can be other transformations that survive the decoupling limit and appear as global symmetries.  First we look for global symmetries of the scalar, i.e., special choices of $\Lambda$ that might survive the decoupling limit to act on $\hat{\phi}$.  We define $\tilde \Lambda={1\over 2}m^2M_{\rm Pl}\Lambda$, so the fields transform as
\begin{align}
\delta \hat h_{\mu\nu}&=\partial_\mu \hat \xi_\nu+\partial_\nu\hat \xi_\mu+{2\over M_{\rm Pl}}{\cal L}_{\hat\xi} \hat h_{\mu\nu}, \nn\\
 \delta \hat A_\mu&={1\over m} \partial_\mu\tilde\Lambda-m\,\hat\xi_\mu+{2\over M_{\rm Pl}}\hat\xi^\nu\partial_\nu \left(\hat A_\mu+{1\over m} \partial_\mu\hat \phi\right), \nn\\ 
 \delta \hat \phi&=- \tilde \Lambda. \label{gaugesyms2d}
 \end{align}
 From this, we see that the first term on the right-hand side of the $ \hat{A}_\mu$ transformation will blow up in the massless decoupling limit with $\tilde{\Lambda}$ held fixed.  One way to avoid this is if $\partial_\mu\tilde\Lambda=0$, i.e., if $\tilde \Lambda$ is a constant.  For this case, the decoupling limit with $\tilde \Lambda$ and $\hat \xi_\mu$ fixed is finite.  The new surviving transformation is the constant shift of $\hat{\phi}$, {which is the global symmetry corresponding to the reducibility parameter of the $U(1)$ gauge symmetry}.
 
The other way to avoid the first term of $\delta \hat A_\mu$ blowing up is if it is cancelled by the second term, i.e., if
\be \label{eq:exact}
\frac{1}{m}\partial_\mu\tilde\Lambda-m \hat \xi_\mu=0 \implies \partial_\mu\tilde\Lambda= \tilde \xi_\mu,
\ee
where we have defined $\tilde \xi_\mu \equiv m^2 \hat{\xi}_\mu$. The remaining field transformations are then given by
\begin{align}
\delta \hat h_{\mu\nu}&={1\over m^2}\left(\partial_\mu \tilde \xi_\nu+\partial_\nu \tilde \xi_\mu\right)+{2\over m^2 M_{\rm Pl}}{\cal L}_{\tilde \xi} \hat h_{\mu\nu}, \nn\\
 \delta \hat A_\mu&={2\over m^2M_{\rm Pl}}\tilde\xi^\nu\partial_\nu \left(\hat A_\mu+{1\over m} \partial_\mu\hat \phi\right), \nn\\ 
 \delta \hat \phi&=- \tilde \Lambda. \label{gaugesyms2e}
 \end{align}
We see that now the first term on the right-hand side of the $ \hat h_{\mu\nu}$ transformation will blow up in the massless decoupling limit with $\tilde \Lambda$ and $\tilde{\xi}_{\mu}$ held fixed unless $\tilde \xi_\mu$ is a Killing vector.  The Killing vectors on Minkowski space are either translations, where $\tilde \xi_\mu=-b_\mu$ for some constant vector $b_\mu$, or Lorentz transformations, where $\tilde \xi_\mu=-m_{\mu\nu}x^\nu$ for some constant antisymmetric tensor $m_{\mu\nu}$.  The Killing vectors corresponding to Lorentz transformations cannot be written as gradients of scalars (since $m_{\mu\nu}x^\nu$ is not closed as a 1-form) so they cannot satisfy the condition \eqref{eq:exact}.  Thus we must use the translation, for which we have the solution $\tilde \Lambda=-b_\mu x^\mu$. 

Under the global symmetry generated by $b_{\mu}$, the fields transform in the following way:
\begin{align}
\delta \hat h_{\mu\nu}&=-{2\over M_{\rm Pl}m^2}b^\rho\partial_\rho \hat h_{\mu\nu}, \nn\\
 \delta \hat A_\mu&=-{2\over M_{\rm Pl}m^2}b^\nu\partial_\nu \hat A_\mu-{2\over M_{\rm Pl}m^3}\partial_\mu\left(b^\nu \partial_\nu\hat \phi\right), \nn\\ 
 \delta \hat \phi&=b_\mu x^\mu. \label{gaugesyms2f}
 \end{align}
These global symmetries are finite for any decoupling limit with the scale $\Lambda_* \leq \Lambda_3$ held fixed. 
The second term on the right-hand side of $\delta \hat A_\mu$ is at the scale $\Lambda_4 \equiv (m^3 M_{\rm Pl})^{1/4}$ and so it appears to blow up if $\Lambda_*>\Lambda_4$, which includes the case $\Lambda_* =\Lambda_3$. However, this term has the form of a $U(1)$ gauge transformation for $A_\mu$ and so it can be cancelled by a transformation \eqref{gaugesyms2b} with 
$\hat \Lambda ={2 \over m^3 M_{\rm Pl}} b^\mu\partial_\mu\hat\phi.$  This leaves
\begin{align}
\delta \hat h_{\mu\nu}&=-{2\over M_{\rm Pl}m^2}b^\rho\partial_\rho \hat h_{\mu\nu}, \nn\\
 \delta \hat A_\mu&=-{2\over M_{\rm Pl}m^2}b^\nu\partial_\nu \hat A_\mu\, ,\nn\\ 
 \delta \hat \phi&=b_\mu x^\mu -{2 \over M_{\rm Pl}m^2} b^\mu\partial_\mu\hat\phi \,.\label{gaugesyms2f}
 \end{align}
The terms carrying the scale $\Lambda_3$ are now simply a translation on $\hat{h}_{\mu \nu}$, $\hat{A}_{\mu}$, and $\hat{\phi}$, and so are part of the ordinary Poincar\'e symmetry in the $\Lambda_3$ decoupling limit. 
 What remains is the galileon symmetry of the decoupling limit acting on the scalar.  We have seen the precise sense in which it appears as the global translation part of the \stu diffeomorphism invariance.  
 Note that the translation terms are absent for any decoupling limit with $\Lambda_* <\Lambda_3$, but the global galileon symmetry is still present. 

We recovered the scalar galileon symmetry in the decoupling limit from the reducibility parameter of the gauge symmetry corresponding to global translations, but what happens to the reducibility parameter of the gauge symmetry corresponding to global Lorentz transformations?  To answer this, we look for global symmetries of the vector, i.e., special choices of $\xi_{\mu}$ that might survive the decoupling limit to act on $A_{\mu}$.  Returning to \eqref{gaugesyms2a}, we see that the fields transform under $\bar \xi_\mu \equiv {1\over 2}M_{\rm Pl}m \xi_\mu$ as
\begin{align}
\delta \hat h_{\mu\nu}&={1\over m}\left(\partial_\mu \bar\xi_\nu+\partial_\nu \bar\xi_\mu\right)+{2\over M_{\rm Pl} m}{\cal L}_{\bar \xi} \hat h_{\mu\nu}\,,\nn\\
 \delta \hat A_\mu&=-\bar\xi_\mu+{2\over M_{\rm Pl}m}\bar\xi^\nu\partial_\nu \left(\hat A_\mu+{1\over m} \partial_\mu\hat \phi\right)\, ,\nn\\
 \delta \hat\phi &=0\,.
 \end{align}
We again see that the first term on the right-hand side of $\delta \hat{h}_{\mu \nu}$ will blow up in the massless decoupling limit with $\bar \xi_\mu$ held fixed unless $\bar \xi_\mu$ is a Killing vector, i.e., $\bar \xi_\mu=-b_\mu$ or $\bar \xi_\mu=-m_{\mu\nu}x^\nu$.  In this case, we are left with
\begin{align}
\delta \hat h_{\mu\nu}&={2\over M_{\rm Pl} m}{\cal L}_{\bar \xi} \hat h_{\mu\nu}\,,\nn\\
 \delta \hat A_\mu&=-\bar\xi_\mu+{2\over M_{\rm Pl}m}{\cal L}_{\bar \xi} \hat A_{\mu}-{2\over M_{\rm Pl}m}\partial_\mu\bar \xi^\nu \hat A_\nu-{2\over M_{\rm Pl}m^2}\partial_\mu\bar\xi^\nu\partial_\nu\hat\phi+{2\over M_{\rm Pl}m^2}\partial_\mu\left(\bar\xi^\nu \partial_\nu\hat\phi\right)\,, \nn\\
 \delta \hat\phi &=0\,.
\label{gaugesyms2g}
 \end{align}
 We have reorganized the terms in the $\delta \hat{A}_\mu$ transformation to make explicit a Lie derivative on $\hat{A}_{\mu}$ in the first term and a total derivative in the last term.  
 
 The total derivative in the last term of the $\delta \hat{A}_\mu$ transformation of \eqref{gaugesyms2g} is a $U(1)$ gauge transformation for $\hat{A}_\mu$ and so it can be cancelled by a transformation \eqref{gaugesyms2b} with gauge parameter
\be
\hat \Lambda =-{2 \over m M_{\rm Pl}} \bar\xi^\nu \partial_\nu\hat\phi=-{2 \over m M_{\rm Pl}}{\cal L}_{\bar \xi} \hat \phi.
\ee 
This leaves
\begin{align}
\delta \hat h_{\mu\nu}&={2\over M_{\rm Pl} m}{\cal L}_{\bar \xi} \hat h_{\mu\nu}\,,\nn\\
 \delta \hat A_\mu&=-\bar\xi_\mu+{2\over M_{\rm Pl}m}{\cal L}_{\bar \xi} \hat A_{\mu}-{2\over M_{\rm Pl}m}\partial_\mu\bar \xi^\nu \hat A_\nu-{2\over M_{\rm Pl}m^2}\partial_\mu\bar\xi^\nu\partial_\nu\hat\phi\,, \nn\\
 \delta \hat\phi &={2 \over m M_{\rm Pl}}{\cal L}_{\bar \xi} \hat \phi\,.
\label{gaugesyms2g0}
 \end{align}
 
 The Lie derivative terms are now simply an ordinary Poincar\'e transformation on $\hat{h}_{\mu \nu}$, $\hat{A}_{\mu}$, and $\hat{\phi}$, and so can be ignored.  In the translation case, $\bar\xi_\mu=-b_\mu$, all the remaining terms in $\delta \hat{A}_{\mu}$ vanish, and so we get nothing new.  In the Lorentz transformation case, $\bar \xi_\mu=-m_{\mu\nu}x^\nu$, we obtain a genuine new global symmetry; in the $\Lambda_3$ decoupling limit, this symmetry comes from the first and final terms in the $\delta \hat{A}_\mu$ transformation of \eqref{gaugesyms2g0}, giving the nonlinear shift symmetry 
 \be \delta \hat{A}_\mu=m_{\mu\nu}\left(x^\nu-{2\over \Lambda_3^3} \partial^\nu\hat \phi\right)\, .\label{fullsymmfinalee}\ee

The transformation \eqref{fullsymmfinalee} is a global symmetry of the decoupling limit of dRGT massive gravity.  Since it only acts on the vector and only mixes in scalars, it is a symmetry of the scalar-vector sector of the theory, as described by \eqref{fulldrgtdecouplesve}. Looking at the first line of \eqref{fulldrgtdecouplesve}, the symmetry completely fixes the structure of the $\hat{F}^2 \hat{\Pi}^p$ terms with $p\geq1$ relative to the photon kinetic term.
 Looking at the second line, the symmetry completely fixes the structure of the $\hat{F}^2 \hat{\Pi}^p$ terms with $p\geq2$ relative to the initial $\hat{F}^2 \hat{\Pi}$ term. And looking at the final line, the symmetry completely fixes the structure of the $\hat{F}^2 \hat{\Pi}^p$ terms with $p\geq3$ relative to the initial $\hat{F}^2 \hat{\Pi}^2$ term. We show how this works in detail in Appendix~\ref{app:decouplingvector}.

Note that the $\hat{\phi}$ term in \eqref{fullsymmfinalee} only survives in the $\Lambda_3$ decoupling limit.  In any decoupling limit with the scale $\Lambda_* <\Lambda_3$ fixed, we would only have the first term in \eqref{fullsymmfinalee}.   Indeed, in this case the vectors in the decoupling limit only appear in the free Maxwell action, for which $\delta A_\mu=m_{\mu\nu}x^\nu$ is a symmetry. 

\subsection{Massless limit of the (A)dS spin-1 theory\label{masslesslimitsecnn}}

We can also see the symmetry \eqref{fullsymmfinalee} by taking the flat space limit of the spin-1 symmetry \eqref{finalshiftsym}.  To take the flat limit, we must introduce a \stu field $\hat{\phi}$ for the longitudinal mode of the vector,
 \be \hat{A}_{\mu}\mapsto\hat{A}_\mu+L\nabla_{\mu}\hat{\phi},\ee
 which introduces a $U(1)$ gauge symmetry,
 \be
\delta \hat{A}_{\mu} = \partial_{\mu} \Lambda, \quad \delta \hat{\phi} = -\frac{1}{L}\Lambda.
 \ee
 The global shift symmetry \eqref{nowvectsymmk2} becomes 
  \be
 \delta \hat{A}_{\mu}=-\frac{2}{\Lambda_2^2}\nabla_\mu\bar{\xi}^{\nu}\,(\hat{A}_{\nu}+L\nabla_{\nu}\hat{\phi})-\bar{\xi}_{\mu} \sqrt{1+{4(\hat{A}_{\mu}+L\nabla_{\mu}\hat{\phi})^2\over {(\Lambda_2^2 L)}^2}}\, . \label{vmlimitintgse}
 \ee
In the flat decoupling limit \cite{deRham:2018svs},
\be L\rightarrow \infty, \quad \Lambda_2 \rightarrow \infty,  \quad{\rm with}~~~ \Lambda_2^2/L ~{\rm \ held~fixed}\, ,\ee
the symmetry \eqref{vmlimitintgse} becomes
 \be
 \delta \hat{A}_{\mu}=-\bar{\xi}_{\mu}-\frac{2L}{\Lambda_2^2}\partial_{\mu}\bar{\xi}^{\nu}\partial_{\nu}\hat{\phi}\, .
 \ee
 In the case where the Killing vector $\bar{\xi}_{\mu}$ is a boost, $-m_{\mu\nu}x^\nu$, we recover \eqref{fullsymmfinalee} after the identification $\Lambda_3\leftrightarrow(\Lambda_2^2/L)^{1/3}$ relevant to this limit \cite{deRham:2018svs}.

\subsection{Pseudo-linear decoupling limit}
\label{sec:abelianvecscal}

We now show how a decoupling limit Lagrangian with an abelian shift symmetry can be derived from the flat space pseudo-linear theory with the potential \eqref{pseudolinearaddse} \cite{Folkerts:2011ev,Hinterbichler:2013eza}.

The flat space \stu prescription for the pseudo-linear theory is the same as for the free theory \cite{Hinterbichler:2013eza},
\be 
h_{\mu\nu}\mapsto h_{\mu\nu}+\frac{1}{m}\left(\partial_\mu A_\nu+\partial_\nu A_\mu\right)+\frac{2}{m^2}\partial_\mu\partial_\nu\phi.
\ee
The gauge symmetry is an abelian version of \eqref{gaugesyms2a},
\begin{align}
\delta h_{\mu\nu}&=\partial_\mu \xi_\nu+\partial_\nu \xi_\mu\, , \nn\\
 \delta A_\mu&=m  \, \partial_\mu\Lambda-m\,\xi_\mu\, , \nn\\ 
 \delta  \phi&=- m^2\Lambda \,. \label{gaugesymspl}
 \end{align}
 
Going through the same arguments as in Section \ref{symfromflatssection}, we can see that the decoupling limit theory will have the same linearized gauge symmetry and galileon shift symmetries as in the non-abelian case.  However, the global vector symmetry will be different. Looking for special choices of $\xi_{\mu}$ that might survive the decoupling limit to act on $A_\mu$, we first write the field transformations in terms of $\bar \xi_\mu\equiv m \xi_\mu$,
\begin{align}
\delta h_{\mu\nu}&={1\over m}\left(\partial_\mu \bar\xi_\nu+\partial_\nu \bar\xi_\mu\right), \nn\\
 \delta A_\mu&=-\bar\xi_\mu. 
 \end{align}
As before, the first term on the right-hand side of $\delta h_{\mu \nu}$ will blow up in the massless decoupling limit with $\bar\xi_\mu$ held fixed unless $\bar \xi_\mu$ is a Killing vector, i.e., unless $\bar \xi_\mu=-b_\mu$ or $\bar \xi_\mu=-m_{\mu\nu}x^\nu$.  In the translation case, when $\bar\xi_\mu=-b_\mu$, this is part of the $U(1)$ gauge symmetry for $A_\mu$, but in the Lorentz transformation case, when $\bar \xi_\mu=-m_{\mu\nu}x^\nu$, the transformation is a genuine global symmetry, 
\be 
\delta A_\mu=m_{\mu\nu}x^\nu,\label{fullsymmfinaleepls2}
\ee
which is just the linear part of \eqref{fullsymmfinalee}.

The scalar-tensor part of the $\Lambda_3$ decoupling limit Lagrangian has a similar form to that of dRGT massive gravity \cite{Hinterbichler:2013eza}, while the scalar-vector part is given by
\be 
\mathcal{L}_{\rm pl}\left({A}, {\phi}\right) =-{1\over 2}{F}_{\mu\nu}^2+\epsilon^{\mu_1\mu_2\mu_3\mu_4} \epsilon^{\nu_1\nu_2\nu_3\nu_4}\left({3 \alpha_3\over 2 \Lambda_3^3 } {F}_{\mu_1\mu_2} {F}_{\nu_1\nu_2} \partial_{\mu_3}\partial_{\nu_3}{\phi} \, \eta_{\mu_4 \nu_4} + \frac{6 \alpha_4}{\Lambda_3^6} {F}_{\mu_1\mu_2} {F}_{\nu_1\nu_2} \partial_{\mu_3}\partial_{\nu_3}{\phi}\, \partial_{\mu_4}\partial_{\nu_4}{\phi} \right)\,, \label{psdecouplepl}
\ee
which has the same form as \eqref{fulldrgtdecouplesve} but without the higher-order interactions. 
The symmetry \eqref{fullsymmfinaleepls2} shifts the Maxwell field strength by a constant, $\delta F_{\mu\nu}\sim m_{\mu\nu}$, and using this and the antisymmetry of the $\epsilon$'s it is straightforward to see that the scalar-vector action \eqref{psdecouplepl} is invariant up to a total derivative. We can also derive \eqref{psdecouplepl} by taking the flat space limit of the ghost-free (A)dS vector theory with the abelian shift symmetry. The symmetry \eqref{fullsymmfinaleepls2} also exists for each vector in the decoupling limit of pseudo-linear interactions involving multiple massless and massive spin-2 fields~\cite{Bonifacio:2019pfg}. 

\subsection{Symmetry algebra}

The generators of the galileon symmetry \eqref{eq:galshift} can be represented as
\begin{align}
C\phi &= 1 \,, & {B^\mu}\phi &= x^\mu \,,\nn\\
CA_\mu &= 0 \,, & {B^\mu}A_\nu &= 0\, .
\end{align}
The standard linearly-realized Poincar\'e generators act as
\begin{align}
 {P_\mu}\phi&=-\partial_\mu\phi ,  & {J_{\mu\nu}}\phi&=(x_\mu\partial_\nu-x_\nu\partial_\mu)\phi, \nn\\
 {P_\mu}A_\nu&=-\partial_\mu A_\nu , & {J_{\mu\nu}}A_\rho&=(x_\mu\partial_\nu-x_\nu\partial_\mu)A_\rho+\left({\cal J}_{\mu\nu}\right)_{\rho}^{\ \sigma}A_{\sigma},
\end{align}
where $\left({\cal J}_{\mu\nu}\right)_{\rho}{}^{\sigma}\equiv\eta_{\mu \rho}\delta_\nu{}^{\sigma}-\eta_{\nu\rho}\delta_\mu{}^{\sigma}$ is the Lorentz generator in the vector representation.
These satisfy the standard commutators of the Poincar\'e algebra,
 \be \left [J_{\mu\nu },P_{\lambda }\right ] =\eta_{\mu\lambda}P_{\nu } - \eta_{\nu\lambda }P_{\mu }, \ \ \ \ \ 
\left [J_{\mu\nu },J_{\lambda\sigma }\right ] = \eta_{\mu\lambda}J_{\nu\sigma }-\eta_{\nu\lambda }J_{\mu\sigma }+\eta_{\nu\sigma}J_{\mu\lambda }-\eta_{\mu\sigma}J_{\nu\lambda }.\ee
The Poincar\'e and galileon generators together close to form the galileon algebra \cite{Goon:2012dy}, whose other non-zero commutators are 
\be
\nn
\left [P_\mu,B_\nu\right ] = \eta_{\mu \nu }C~,\ \ \ 
\left[J_{\mu\nu},B_{\lambda}\right] = \eta_{\mu\lambda }B_{\nu }-\eta_{\nu\lambda}B_{\mu}~.
\ee
 
There is now an additional antisymmetric traceless generator associated with the new symmetry \eqref{fullsymmfinalee},
\begin{equation}
M_{\lambda\nu}A_{\mu}=\eta_{\mu[\lambda}x_{\nu]}-{2\over \Lambda_3^3}\eta_{\mu[\lambda}\partial_{\nu]} \phi\, ,\  \ \ M_{\lambda\nu}\phi_{}=0. \label{newsymmmforme}
\end{equation}
The only non-zero commutator of this new symmetry with itself or with the remaining generators of the galileon algebra is the one dictated by Lorentz invariance,
\be
\left [J_{\mu\nu },M_{\lambda\sigma }\right ] = \eta_{\mu\lambda}M_{\nu\sigma }-\eta_{\nu\lambda }M_{\mu\sigma }+\eta_{\nu\sigma}M_{\mu\lambda }-\eta_{\mu\sigma}M_{\nu\lambda }. 
\ee
In particular, all commutators of the new symmetry with the translations and galileon transformations vanish.  In saying that these commutators vanish, we really mean that they vanish up to a $U(1)$ gauge transformation on the photon.  For example, the commutator of the new symmetry with a translation by $a^{\mu}$ acting on the vector field is
\be m^{\nu\lambda}a^\sigma\left[M_{\nu\lambda},P_\sigma\right]A_{\mu}=\partial_\mu\left(m^{\nu\lambda}a_\nu x_\lambda\right) \,,\label{gaugeideale}\ee
and the commutator of the new symmetry with a galileon shift by $b^{\mu}$ acting on the vector field is
\be m^{\nu\lambda}b^\sigma\left[M_{\nu\lambda},B_\sigma\right]A_{\mu}=\partial_\mu\left({2\over \Lambda_3^3}m^{\nu\lambda}b_\nu x_\lambda\right) \,.\label{gaugeideale}\ee
When a system possesses both gauge and global symmetries, as this one does, the global symmetry algebra is only defined modulo the ideal of gauge symmetries \cite{Barnich:2001jy}.

The global symmetry algebra does not depend on the terms proportional to ${1/ \Lambda_3^3}$ in \eqref{newsymmmforme}, so it is the same algebra as that of the pseudo-linear theory with $ \delta  A_\mu=m_{\mu\nu}x^\nu$.  It might be thought that theories nonlinearly realizing the same global symmetry algebra must be the same up to field redefinitions, which would mean that the pseudo-linear decoupling limit is secretly the same theory as the dRGT decoupling limit.  However, as can be checked by explicitly computing four-point scattering amplitudes, these theories are not the same.  One way to understand why these theories are inequivalent is to note that the structure of the full algebra including the gauge symmetries is different, as can be seen from~\eqref{gaugeideale}, and the coset construction of theories with gauge fields is sensitive to the structure of the gauge algebra~\cite{Ivanov:1976zq,Goon:2014ika,Goon:2014paa}.

\section{Conclusions}
\label{sec:concl}

In this paper we have studied interacting theories of massive spin-1 fields on (A)dS space with the special mass given by
\be
m^2={6\over L^2},
\ee
corresponding to fields that have a symmetry under a shift by an (A)dS Killing vector (this is a $k=0$ shift symmetry in the classification of Ref.~\cite{Bonifacio:2018zex}).  The interactions are governed either by the abelian symmetry algebra of the free theory or by the non-abelian algebra $\mathfrak{so}(5)\oplus \mathfrak{so}(5)$.  The theories with the non-abelian shift symmetry include the spin-1 theories that arise in the (A)dS decoupling limit of massive gravity \cite{deRham:2018svs}.

Nonlinear massive spin-1 theories have been of interest recently \cite{Tasinato:2014eka,Heisenberg:2014rta,Hull:2014bga,Tasinato:2014mia,Allys:2015sht,Hull:2015uwa,Charmchi:2015ggf,Jimenez:2016isa,Allys:2016jaq,Heisenberg:2017mzp,ErrastiDiez:2019ttn,ErrastiDiez:2019trb,GallegoCadavid:2019zke,Jimenez:2019hpl}.  These theories are typically constructed to be ghost-free, so that they propagate nonlinearly only the expected three degrees of freedom of a massive spin-1 particle.  In the massless decoupling limit, they reduce to galileons or generalizations thereof, which can possess enhanced shift symmetries.  However, these decoupling limit shift symmetries are generally not present in the full theory away from the decoupling limit.  Here we have found (A)dS spin-1 theories that truly have a galileon-like shift symmetry away from any such limits.

We discussed how these theories are constructed from invariant building blocks using the decoupling limit of massive gravity in (A)dS space.  It should also be possible, though more complicated, to construct these building blocks in a more direct way from the coset construction for the symmetry breaking pattern, as was done for the galileons in Ref.~\cite{Goon:2012dy}.  This would allow one to search for possible Wess--Zumino terms which are not constructed from the invariant building blocks.  These terms, if they exist, would be missed by our construction.

Interacting theories of scalars in (A)dS space with $k\leq 2$ shift symmetries were studied in Ref.~\cite{Bonifacio:2018zex}, which are generalizations of the galileon and special galileon to (A)dS space.  Here we have shown that there also exist interacting vector theories with a non-abelian  $k=0$ shift symmetry. To further explore the space of interacting (A)dS field theories with shift symmetries, it would be nice to have a classification of non-abelian algebras containing the (A)dS isometry algebra as a sub-algebra and to find interactions invariant under the corresponding symmetry transformations. Some example algebras are the finite higher-spin algebras found in Ref.~\cite{Joung:2015jza}. For example, by truncating the generators of one of their algebras to the even-spin Killing tensors, one obtains an algebra with generators parametrized by ambient space tensors that have the symmetries of the following traceless Young diagrams:
\be
\left\{ \, { \Yvcentermath1 \yng(1,1)} \ , \, { \Yvcentermath1 \yng(3,1)}^{ \, T} \right\} \, .
\ee
This algebra is $\mathfrak{so}\left( 14\right)$ in (A)dS$_4$\footnote{In AdS$_D$ it is $\mathfrak{so}\left( \frac{1}{2}D(D+3)\right)$. We thank Euihun Joung and Karapet Mkrtchyan for pointing this out.} and it has the correct generators to be the nonlinearly realized global symmetry of a vector theory in (A)dS space with a non-abelian $k=2$ shift symmetry, which corresponds to a vector with mass squared $m^2={20/ L^2}$.  Similarly, the flat space contraction of this algebra could govern a scalar-vector theory in flat space with a large global symmetry. It would be interesting to search for these theories.

We have considered $k=0$ vector symmetries, but scalar-vector theories with abelian shift symmetries with higher values of $k$ will also arise from the decoupling limits of massive higher-spin theories, as discussed in Ref.~\cite{Bellazzini:2019bzh}.  For example, the massive spin-3 scalar-vector decoupling limit interactions of Ref.~\cite{Bellazzini:2019bzh} have a non-trivial abelian $k=1$ vector shift symmetry. This should generalize to arbitrary $k$ using the decoupling limits of massive higher-spins that interact via special potentials that generalize the pseudo-linear interactions \cite{Bonifacio:2017iry,Bonifacio:2018vzv, Bellazzini:2019bzh}.  In Appendix~\ref{app:higherkvects} we write down these scalar-vector interactions explicitly for all even values of $k$. 

At the quantum level, the galileons and other shift-symmetric theories have terms which satisfy non-renormalization theorems \cite{Luty:2003vm,Nicolis:2004qq,Hinterbichler:2010xn,Goon:2016ihr}.  It would be interesting to study quantum corrections on (A)dS space for theories with shift symmetries to see how the symmetries constrain quantum corrections. Some work in this direction for the shift-symmetric scalars can be found in Ref.~\cite{Folacci:1992xc} and the de Sitter quantization of the shift-symmetric vectors has been studied in Ref.~\cite{Miao:2010vs}.  Another direction to explore is whether gauging the new global symmetries we have found can help to recover massive gravity from its flat space decoupling limit \cite{Garcia-Saenz:2019yok}. 

Finally, as mentioned in the introduction, there is a close connection between shift symmetries of massless particles in flat space and soft limits of scattering amplitudes. Consider an amplitude $\mathcal{A}$ where the $i$\textsuperscript{th} particle, $\psi$, has spin $s_i$ and momentum $p_i$. Then taking the soft limit of particle $i$, we have
\be
\lim_{p_i \rightarrow 0} \mathcal{A} = \mathcal{O} \left(p_i^{\sigma_{\psi}+s_i} \right),
\ee
where $\sigma_{\psi}$ is an integer whose definition agrees in four dimensions with the holomorphic soft weight of Ref.~\cite{Elvang:2018dco}. For the flat space scalar-vector theories we have discussed, the naive expectation---based on the form of their symmetry transformations---is that their amplitudes should have $\sigma_{\phi} = 2$ and $\sigma_A=1$. However, by explicit calculation we find that at five points and above this is generically \textit{not} the case.  Thus we expect that the Ward identity corresponding to the non-linear symmetry \eqref{vectsyminpe} does not always lead to a simple vanishing in the soft limit, but rather a soft theorem relating soft limits and other amplitudes.
An example that does have the expected vanishing soft behavior at six points, and hence should be constructible by soft recursion~\cite{Cheung:2015ota, Cheung:2016drk, Elvang:2018dco}, is the theory with the abelian shift symmetry and no cubic interaction. 

\vspace{-.4cm}
\paragraph{Acknowledgements:} We would like to thank Claudia de Rham, Euihun Joung, Karapet Mkrtchyan, Shruti Paranjape, Diederik Roest, and David Stefanyszyn for helpful conversations and comments, and we especially thank Rachel Rosen for early collaboration as well as conversations and input throughout the work.  We would like to thank Daniel Baumann, the Delta-ITP, and the University of Amsterdam for arranging a long-term visit during which this work was largely completed.
KH acknowledges support from DOE grant DE-SC0019143.   
AJ is supported by NASA grant NNX16AB27G.  

\appendix

\section{Vector symmetry from projection}
\label{projapp}
In this appendix we derive the form of the \stu diffeomorphism symmetry acting on the massive vector in (A)dS space, using the notation of Appendix C of Ref.~\cite{deRham:2018svs}. Briefly, we consider embedding (A)dS space in a Minkowski space of one higher dimension which has flat coordinates $Z^A$. The embedding is defined via the coordinates $X^A = (Y,x^\mu)$. The $Y$ coordinate labels the radius of the (A)dS hyperboloid, which defines a foliation of the 5-dimensional Minkowksi space by (A)dS slices. The particular (A)dS slice we are interested in sits at $Y=0$ and has radius $L$ or $H^{-1}$, depending on the signature.

Before projecting to the (A)dS surface, the \stu action of massive gravity in flat ambient space has the diffeomorphism gauge symmetry
\begin{align}
\delta h_{AB}&=\partial_A\xi_B+\partial_B\xi_A+\mathcal{L}_{\xi}h_{AB},\\
\delta V_{A}&=-\xi_{A}+\xi^{C}\partial_{C}V_{A}.
\end{align}
Here $\xi^A(Z)$ is an ambient diffeomorphism parameter, which should preserve the (A)dS surface, so it must satisfy
\be Z_A \xi^A\big|_{Y=0}=0.\ee
Projecting to the (A)dS surface gives
\begin{align}
\delta h_{\mu\nu}&=(\nabla_A\xi_B+\nabla_B\xi_A+\mathcal{L}_{\xi}h_{AB})\left.\frac{\partial Z^A}{\partial x^{\mu}}\frac{\partial Z^B}{\partial x^{\nu}}\right\rvert_{Y=0}\\
&=(\nabla_M\xi_N+\nabla_N\xi_M+\mathcal{L}_{\xi}h_{MN})\left.\frac{\partial X^M}{\partial x^{\mu}}\frac{\partial X^N}{\partial x^{\nu}}\right\rvert_{Y=0}\, ,\\
\delta A_{\mu}&=(-\xi_{A}+\xi^{C}\nabla_{C}V_{A})\left.\frac{\partial Z^A}{\partial x^{\mu}}\right\rvert_{Y=0}\\
&=(-\xi_{M}+\xi^{P}\nabla_{P}V_{M})\left.\frac{\partial X^M}{\partial x^{\mu}}\right\rvert_{Y=0} \\
&=-\xi_{\mu}+\xi^{\rho}\nabla_{\rho}A_{\mu}+\frac{1}{L}\xi_{\mu}V^Y\,,
\end{align}
where 
\begin{align}
V^Y=L\left(1-\sqrt{1+{1\over L^2}A^2}\right) \quad {\rm when}~~Y=0\, ,
\end{align}
which leads to the symmetry given in \eqref{diffbeforenorme},
\begin{align}
 \delta A_{\mu} =-\xi_{\mu}+\xi^{\rho}\nabla_{\rho}A_{\mu}+\xi_{\mu}\left(1-\sqrt{1+{1\over L^2}A^2}\right).
\end{align}

\section{Massive gravity decoupling limit interactions from symmetry}
\label{app:decouplingvector}
In this appendix, we show how the scalar-vector interactions in the flat space decoupling limit of dRGT massive gravity are fixed in terms of a finite number of seed interactions by demanding invariance under the Lorentz-like symmetry~\eqref{vectsyminpe}. 

Consider splitting a transformation of the form~\eqref{vectsyminpe} as
\begin{align} 
\delta A_{\mu} = \delta^{(0)} A_{\mu} + \delta^{(1)} A_{\mu} = m_{\mu \nu} x^{\nu} + 2 \alpha m_{\mu \nu} \partial^{\nu} \phi, \label{eq:Asymmetry}
\end{align}
where $\alpha$ is a free dimensionful parameter.
The field strength transforms as
\begin{align} \label{eq:Ftransform}
\delta^{(0)} F_{\mu \nu} = -2 m_{\mu \nu}, \quad\qquad \delta^{(1)} F_{\mu \nu} = 4\alpha m_{\lambda [\mu} \Pi_{\nu]}{}^{\lambda} \,,
\end{align}
where $\Pi_{\mu \nu} \equiv \partial_{\mu} \partial_{\nu} \phi$.
We want to find invariant interactions $\mathcal{L}_n$, which we can expand as
\be \label{eq:Ln}
\mathcal{L}_n =\mathcal{L}^{(n)}_n+\mathcal{L}^{(n+1)}_n+\mathcal{L}^{(n+2)}_n+ \cdots,
\ee
where  $n=2, 3, \ldots$ and the superscript denotes the order in fields. In order for this to be invariant we require that the following conditions are satisfied:
\begin{align}
\delta^{(0)} \mathcal{L}^{(n)}_n & = 0, \label{eq:initial}\\
\delta^{(0)} \mathcal{L}^{(k+1)}_n +\delta^{(1)} \mathcal{L}^{(k)}_n& = 0, \quad k=n, n+1, \ldots \, . \label{eq:subsequent}
\end{align}
We restrict our attention to the scalar-vector interactions that have the fewest derivatives per field, since these are the interactions that appear in the decoupling limit of dRGT massive gravity. We expect these to correspond to the most general invariant ghost-free interactions involving a single scalar and vector.

Solving the initial constraint \eqref{eq:initial} is straightforward; the solution is given by
\be \label{eq:abelian}
\mathcal{L}_n^{(n)} = c_n \alpha^{n-2} \eta^{\mu_1 \nu_1 \ldots \mu_n \nu_n} F_{\mu_1 \mu_2} F_{\nu_1 \nu_2} \Pi_{\mu_3 \nu_3} \cdots \Pi_{\mu_n \nu_n},
\ee
where 
\be
\eta^{\mu_1 \nu_1 \cdots \mu_n \nu_n} = -\frac{1}{(m-n)!}\epsilon^{\mu_1\cdots\mu_n\alpha_{n+1}\cdots\alpha_{m}} \epsilon^{\nu_1\cdots\nu_n}_{~~~~~~~\alpha_{n+1}\cdots\alpha_{m}}
\ee
is the generalized Kronecker delta and $c_n$ is a dimensionless constant. These are just the interactions discussed in Section~\ref{sec:abelianvecscal}. When we try to find $\mathcal{L}_n^{(k+1)}$ in terms of $\mathcal{L}_n^{(k)}$ using \eqref{eq:subsequent}, 
there is an ambiguity since we can always add the abelian solution $\mathcal{L}_{k+1}^{(k+1)}$ to  $\mathcal{L}_n^{(k+1)}$ with a free parameter. To uniquely determine the whole tower of interactions in terms of an initial seed we thus need to give a prescription for removing this ambiguity, i.e., for fixing the homogeneous term. One choice---which seems to minimize the number of terms---is that we choose $\mathcal{L}_n^{(k)}$ with $k>n$ to not contain the term $[F^2] [ \Pi ]^{k-2}$, where $(F^2)^{\mu_1 \mu_2} \equiv F^{\mu_1}{}_{\lambda}F^{\mu_2 \lambda}$ and $[\cdot]$ denotes the trace. This can always be achieved by adding a suitable multiple of the abelian term and results in no loss of generality of the total Lagrangian $\sum_n \mathcal{L}_n$. For example, for $n=2$ with $c_2 =-1/4 $ this gives
\begin{align}
\mathcal{L}_2^{(2)} & = -\frac{1}{2} [F^2], \\
\mathcal{L}_2^{(3)} & = \alpha (F^2)^{\mu_1 \mu_2} \Pi_{\mu_1 \mu_2}, \\
\mathcal{L}_2^{(4)} & = -\alpha^2 (F^2)^{\mu_1 \mu_2} \Pi^2_{\mu_1 \mu_2}-\alpha^2 F^{\mu_1 \mu_2} F^{\nu_1 \nu_2} \Pi_{\mu_1 \nu_1} \Pi_{\mu_2 \nu_2}, \\
\mathcal{L}_2^{(5)} & = \alpha^3 (F^2)^{\mu_1 \mu_2} \Pi^3_{\mu_1 \mu_2}+3\alpha^3   F^{\mu_1 \mu_2} F^{\nu_1 \nu_2} \Pi^2_{\mu_1 \nu_1} \Pi_{\mu_2 \nu_2}, \\
\mathcal{L}_2^{(6)} & = -\alpha^4 (F^2)^{\mu_1 \mu_2} \Pi^4_{\mu_1 \mu_2}-4\alpha^4 F^{\mu_1 \mu_2} F^{\nu_1 \nu_2} \Pi^3_{\mu_1 \nu_1} \Pi_{\mu_2 \nu_2}-3\alpha^4 F^{\mu_1 \mu_2} F^{\nu_1 \nu_2} \Pi^2_{\mu_1 \nu_1} \Pi^2_{\mu_2 \nu_2}, \\
\mathcal{L}_2^{(7)} & = \alpha^5 (F^2)^{\mu_1 \mu_2} \Pi^5_{\mu_1 \mu_2}+5\alpha^5 F^{\mu_1 \mu_2} F^{\nu_1 \nu_2} \Pi^4_{\mu_1 \nu_1} \Pi_{\mu_2 \nu_2}+10\alpha^5 F^{\mu_1 \mu_2} F^{\nu_1 \nu_2} \Pi^3_{\mu_1 \nu_1} \Pi^2_{\mu_2 \nu_2}, \\
 &~\vdots \nonumber
\end{align}
where powers of $\Pi$ are defined in the obvious way: $\Pi^n_{\mu \nu} \equiv \Pi_{\mu}{}^{\lambda_1}\Pi_{\lambda_1}{}^{\lambda_2}\cdots\Pi_{\lambda_n\nu}$.

It is natural to search for a more elegant approach to solving Eq.~\eqref{eq:subsequent} as opposed to brute force solving with a general ansatz. Looking at the transformation of the field strength \eqref{eq:Ftransform}, we might guess that we can write $\mathcal{L}_n^{(k+1)}$ in terms of a variation of $\mathcal{L}_n^{(k)}$. It turns out that the following relation gives a solution:\footnote{Plugging this into Eq.~\eqref{eq:subsequent}, we find that it is satisfied only if
\be \label{eq:relation}
 m_{\nu_1 \nu_2} \frac{\delta^2 \mathcal{L}_n^{(k)}}{\delta F_{\nu_1 \nu_2} \delta F_{\mu_1 \mu_2}} F_{\lambda [\mu_1} \Pi_{\mu_2]}{}^{\lambda} = \frac{\delta \mathcal{L}^{(k)}_n}{\delta F_{\mu_1 \mu_2}} m_{\lambda [\mu_1} \Pi_{\mu_2]}{}^{\lambda} \, .
\ee
This does not hold for general interactions $\mathcal{L}_n^{(k)}$, but we can check that it is true for those of the form $F_{\mu_1 \mu_2} F_{\nu_1 \nu_2} T^{\mu_1 \mu_2 \nu_1 \nu_2}(\eta, \Pi)$, where $T^{\mu_1 \mu_2 \nu_1 \nu_2}$ is any tensor built from $\eta_{\mu \nu}$'s and $\Pi_{\mu \nu}$'s. The abelian interactions are of this form and, if $\mathcal{L}_n^{(k)}$ is of this form, \eqref{eq:recursive} ensures that $\mathcal{L}_n^{(k+1)}$ is as well.}
\be \label{eq:recursive}
\mathcal{L}_n^{(k+1)} = \alpha F_{\lambda [\mu} \Pi_{\nu]}{}^{\lambda}  \frac{\delta \mathcal{L}_n^{(k)}}{\delta F_{\mu \nu}} \,.
\ee
The higher-order terms generated by \eqref{eq:recursive} do not contain the trace $[F^2]$, so this fixes the abelian term ambiguity in the same way as the prescription mentioned above. 

Using Eq.~\eqref{eq:recursive} recursively, we can build up $\mathcal{L}_n$ order-by-order in terms of the initial seed $\mathcal{L}_n^{(n)}$. We can thus formally write a closed-form solution as
\be
\mathcal{L}_n =\frac{1}{1- \alpha \mathcal{D}} \mathcal{L}_n^{(n)} \equiv \sum_{i=0}^{\infty} \alpha^i \mathcal{D}^i  \mathcal{L}_n^{(n)}, \quad {\rm where} \quad \mathcal{D} \equiv  F_{\lambda [\mu} \Pi_{\nu]}{}^{\lambda}\frac{\delta}{\delta F_{\mu \nu}}.
\ee
We can be a bit more explicit and write this as
\be
\mathcal{L}_n =  \sum_{i=0}^{\infty}\sum_{j=0}^i c_n \alpha^{n+i-2} {{i}\choose{j}} S^{(i-j+1)}_{\mu_1 \mu_2} S^{(j+1)}_{\nu_1 \nu_2} \eta^{\mu_1 \nu_1 \ldots \mu_n \nu_n}\Pi_{\mu_3 \nu_3} \cdots \Pi_{\mu_n \nu_n},
\ee
where we have defined
\be
S^{(n+1)}_{\mu_1 \nu_1} \equiv \mathcal{D}^{n} F_{\mu_1 \nu_1} = F_{\lambda [\mu_n} \Pi^{\lambda}{}_{\nu_n]} \delta^{\nu_n}{}_{[\mu_{n-1}} \Pi^{\mu_n}{}_{\nu_{n-1}]}\delta^{\nu_{n-1}}{}_{[\mu_{n-2}} \Pi^{\mu_{n-1}}{}_{\nu_{n-2}]}   \cdots \delta^{\nu_2}{}_{[\mu_{1}} \Pi^{\mu_2}{}_{\nu_{1}]} \, .
\ee
In $D$ dimensions we can thus write the general invariant interaction as
\be \label{eq:general}
\mathcal{L} = \sum_{n=2}^{D} \mathcal{L}_n =  \sum_{n=2}^{D}  \sum_{i=0}^{\infty} \sum_{j=0}^i c_n \alpha^{n+i-2} {{i}\choose{j}} S^{(i-j+1)}_{\mu_1 \mu_2} S^{(j+1)}_{\nu_1 \nu_2} \eta^{\mu_1 \nu_1 \ldots \mu_n \nu_n}\Pi_{\mu_3 \nu_3} \cdots \Pi_{\mu_n \nu_n}.
\ee

The full scalar-vector decoupling limit interactions of massive gravity were first derived in Refs.~\cite{Gabadadze:2013ria,Ondo:2013wka}. In four dimensions with the scalar kinetic term normalized as $-6 (\partial \phi)^2$, they are invariant under the symmetry \eqref{eq:Asymmetry} with
\be \label{eq:alphavalue}
\alpha = - \frac{1}{\Lambda_3^3}.
\ee
Using the relation \eqref{eq:recursive} to generate the higher-order terms, these decoupling limit interactions are generated by the following seeds:
\begin{align}
\mathcal{L}^{(2)}_2 & = -\frac{1}{2} F_{\mu_1 \mu_2}F^{\mu_1 \mu_2}, \\
\mathcal{L}^{(3)}_3 & = \frac{3 \alpha_3 +1}{2\Lambda_3^3} \epsilon^{\mu_1 \mu_2 \mu_3 \lambda} \epsilon^{\nu_1 \nu_2 \nu_3}{}{}{}_{\lambda} F_{\mu_1 \mu_2} F_{\nu_1 \nu_2} \Pi_{\mu_3 \nu_3}, \\
\mathcal{L}^{(4)}_4 & = \frac{12\alpha_4+3 \alpha_3}{2\Lambda_3^6}\epsilon^{\mu_1 \mu_2 \mu_3 \mu_4} \epsilon^{\nu_1 \nu_2 \nu_3 \nu_4} F_{\mu_1 \mu_2} F_{\nu_1 \nu_2} \Pi_{\mu_3 \nu_3}  \Pi_{\mu_4 \nu_4}.
\end{align}
Or, equivalently, the interactions are given by Eq.~\eqref{eq:general} with $D=4$, $\alpha$ given by Eq.~\eqref{eq:alphavalue}, and
\be
c_2 = -\frac{1}{4}, \quad c_3 = \frac{1}{2} \left( 3\alpha_3+1\right), \quad c_4 = -\frac{3}{2} \left( 4 \alpha_4+\alpha_3 \right) \, .
\ee

\section{Higher-$k$ abelian shift symmetries}
\label{app:higherkvects}
The existence of special potentials for massive higher-spin fields \cite{Bonifacio:2017iry,Bonifacio:2018vzv, Bellazzini:2019bzh} suggests that there should exist scalar-vector interactions with nontrivial higher-$k$ abelian shift symmetries, which would appear in the decoupling limit. These interactions would have the following abelian vector symmetry:
\be
\delta A_\mu=m_{\mu\nu_1,\,\nu_2\cdots\nu_{k+1}}x^{\nu_1}\cdots x^{\nu_{k+1}},
\ee
where $m_{\mu\nu_1,\,\nu_2\cdots\nu_{k+1}}$ is a mixed symmetry constant tensor that is antisymmetric in its first two indices and completely traceless.

We can look directly for interactions possessing these symmetries, using the expected form of the higher-spin interactions to determine how many derivatives should appear. This leads to the following $n$-point interactions with an order-$k$ vector shift symmetry and a (trivial) order-$(k+1)$ scalar shift symmetry, for every $n\geq 3$ and even $k\geq0$:
\begin{align}
\mathcal{L}_{k,n}(A, \phi) &=\left( \prod_{i=1}^{\frac{k+2}{2}} \eta^{\mu_1^{(i)} \nu_1^{(i)} \cdots \mu_n^{(i)} \nu_n^{(i)}}\right) \left( \prod_{j=2}^{\frac{k+2}{2}} \eta_{\mu_2^{(j)} \nu_2^{(j)}}\right) \partial_{\mu_1^{(2)}}\cdots\partial_{\mu_1^{\left(k/2+1\right)}} F_{\mu_1^{(1)} \mu_2^{(1)}} \partial_{\nu_1^{(2)}}\cdots\partial_{\nu_1^{\left(k/2+1\right)}} F_{\nu_1^{(1)} \nu_2^{(1)}} \nonumber \\
&~~~~~~~ \times \prod_{l=3}^n \partial_{\mu_l^{(1)}} \partial_{\nu_l^{(1)}} \cdots \partial_{\mu_l^{(k/2+1)}} \partial_{\nu_l^{(k/2+1)}} \phi \, .
\end{align}
These should correspond to part of the decoupling limit of an interacting massive spin-$(k+2)$ particle with interactions chosen to maximally improve the high-energy growth of scattering amplitudes. For odd spins the corresponding scalar-vector interactions exist only for even $n$ and are not uniquely fixed by the symmetry, so we do not consider them here.

\renewcommand{\em}{}
\bibliographystyle{utphys}
\addcontentsline{toc}{section}{References}
\bibliography{vectors-arxiv}

\providecommand{\href}[2]{#2}\begingroup\raggedright\begin{thebibliography}{10}

\bibitem{Adler:1964um}
S.~L. Adler, ``{Consistency conditions on the strong interactions implied by a
  partially conserved axial vector current},''
  \href{http://dx.doi.org/10.1103/PhysRev.137.B1022}{{\em Phys. Rev.} {\bf 137}
  (1965)  B1022--B1033}.
[,140(1964)].

\bibitem{Adler:1965ga}
S.~L. Adler, ``{Consistency conditions on the strong interactions implied by a
  partially conserved axial-vector current. II},''
  \href{http://dx.doi.org/10.1103/PhysRev.139.B1638}{{\em Phys. Rev.} {\bf 139}
  (1965)  B1638--B1643}.
[,152(1965)].

\bibitem{Nicolis:2008in}
A.~Nicolis, R.~Rattazzi, and E.~Trincherini, ``{The Galileon as a local
  modification of gravity},''
  \href{http://dx.doi.org/10.1103/PhysRevD.79.064036}{{\em Phys. Rev.} {\bf
  D79} (2009)  064036},
\href{http://arxiv.org/abs/0811.2197}{{\tt arXiv:0811.2197 [hep-th]}}.

\bibitem{Trodden:2011xh}
M.~Trodden and K.~Hinterbichler, ``{Generalizing Galileons},''
  \href{http://dx.doi.org/10.1088/0264-9381/28/20/204003}{{\em Class. Quant.
  Grav.} {\bf 28} (2011)  204003},
\href{http://arxiv.org/abs/1104.2088}{{\tt arXiv:1104.2088 [hep-th]}}.

\bibitem{Hinterbichler:2014cwa}
K.~Hinterbichler and A.~Joyce, ``{Goldstones with Extended Shift Symmetries},''
  \href{http://dx.doi.org/10.1142/S0218271814430019}{{\em Int. J. Mod. Phys.}
  {\bf D23} (2014) no.~13, 1443001},
\href{http://arxiv.org/abs/1404.4047}{{\tt arXiv:1404.4047 [hep-th]}}.

\bibitem{Griffin:2014bta}
T.~Griffin, K.~T. Grosvenor, P.~Horava, and Z.~Yan, ``{Scalar Field Theories
  with Polynomial Shift Symmetries},''
  \href{http://dx.doi.org/10.1007/s00220-015-2461-2}{{\em Commun. Math. Phys.}
  {\bf 340} (2015) no.~3, 985--1048},
\href{http://arxiv.org/abs/1412.1046}{{\tt arXiv:1412.1046 [hep-th]}}.

\bibitem{deRham:2010eu}
C.~de~Rham and A.~J. Tolley, ``{DBI and the Galileon reunited},''
  \href{http://dx.doi.org/10.1088/1475-7516/2010/05/015}{{\em JCAP} {\bf 1005}
  (2010)  015},
\href{http://arxiv.org/abs/1003.5917}{{\tt arXiv:1003.5917 [hep-th]}}.

\bibitem{Hinterbichler:2010xn}
K.~Hinterbichler, M.~Trodden, and D.~Wesley, ``{Multi-field galileons and
  higher co-dimension branes},''
  \href{http://dx.doi.org/10.1103/PhysRevD.82.124018}{{\em Phys. Rev.} {\bf
  D82} (2010)  124018},
\href{http://arxiv.org/abs/1008.1305}{{\tt arXiv:1008.1305 [hep-th]}}.

\bibitem{Goon:2011qf}
G.~Goon, K.~Hinterbichler, and M.~Trodden, ``{Symmetries for Galileons and DBI
  scalars on curved space},''
  \href{http://dx.doi.org/10.1088/1475-7516/2011/07/017}{{\em JCAP} {\bf 1107}
  (2011)  017},
\href{http://arxiv.org/abs/1103.5745}{{\tt arXiv:1103.5745 [hep-th]}}.

\bibitem{Goon:2011uw}
G.~Goon, K.~Hinterbichler, and M.~Trodden, ``{A New Class of Effective Field
  Theories from Embedded Branes},''
  \href{http://dx.doi.org/10.1103/PhysRevLett.106.231102}{{\em Phys. Rev.
  Lett.} {\bf 106} (2011)  231102},
\href{http://arxiv.org/abs/1103.6029}{{\tt arXiv:1103.6029 [hep-th]}}.

\bibitem{Goon:2012dy}
G.~Goon, K.~Hinterbichler, A.~Joyce, and M.~Trodden, ``{Galileons as
  Wess-Zumino Terms},'' \href{http://dx.doi.org/10.1007/JHEP06(2012)004}{{\em
  JHEP} {\bf 06} (2012)  004},
\href{http://arxiv.org/abs/1203.3191}{{\tt arXiv:1203.3191 [hep-th]}}.

\bibitem{Cheung:2014dqa}
C.~Cheung, K.~Kampf, J.~Novotny, and J.~Trnka, ``{Effective Field Theories from
  Soft Limits of Scattering Amplitudes},''
  \href{http://dx.doi.org/10.1103/PhysRevLett.114.221602}{{\em Phys. Rev.
  Lett.} {\bf 114} (2015) no.~22, 221602},
\href{http://arxiv.org/abs/1412.4095}{{\tt arXiv:1412.4095 [hep-th]}}.

\bibitem{Cheung:2015ota}
C.~Cheung, K.~Kampf, J.~Novotny, C.-H. Shen, and J.~Trnka, ``{On-Shell
  Recursion Relations for Effective Field Theories},''
  \href{http://dx.doi.org/10.1103/PhysRevLett.116.041601}{{\em Phys. Rev.
  Lett.} {\bf 116} (2016) no.~4, 041601},
\href{http://arxiv.org/abs/1509.03309}{{\tt arXiv:1509.03309 [hep-th]}}.

\bibitem{Cheung:2016drk}
C.~Cheung, K.~Kampf, J.~Novotny, C.-H. Shen, and J.~Trnka, ``{A Periodic Table
  of Effective Field Theories},''
  \href{http://dx.doi.org/10.1007/JHEP02(2017)020}{{\em JHEP} {\bf 02} (2017)
  020},
\href{http://arxiv.org/abs/1611.03137}{{\tt arXiv:1611.03137 [hep-th]}}.

\bibitem{Cheung:2018oki}
C.~Cheung, K.~Kampf, J.~Novotny, C.-H. Shen, J.~Trnka, and C.~Wen, ``{Vector
  Effective Field Theories from Soft Limits},''
  \href{http://dx.doi.org/10.1103/PhysRevLett.120.261602}{{\em Phys. Rev.
  Lett.} {\bf 120} (2018) no.~26, 261602},
\href{http://arxiv.org/abs/1801.01496}{{\tt arXiv:1801.01496 [hep-th]}}.

\bibitem{Klein:2017npd}
R.~Klein, D.~Roest, and D.~Stefanyszyn, ``{Spontaneously Broken Spacetime
  Symmetries and the Role of Inessential Goldstones},''
  \href{http://dx.doi.org/10.1007/JHEP10(2017)051}{{\em JHEP} {\bf 10} (2017)
  051},
\href{http://arxiv.org/abs/1709.03525}{{\tt arXiv:1709.03525 [hep-th]}}.

\bibitem{Klein:2018ylk}
R.~Klein, E.~Malek, D.~Roest, and D.~Stefanyszyn, ``{No-go theorem for a gauge
  vector as a spacetime Goldstone mode},''
  \href{http://dx.doi.org/10.1103/PhysRevD.98.065001}{{\em Phys. Rev.} {\bf
  D98} (2018) no.~6, 065001},
\href{http://arxiv.org/abs/1806.06862}{{\tt arXiv:1806.06862 [hep-th]}}.

\bibitem{Bogers:2018zeg}
M.~P. Bogers and T.~Brauner, ``{Lie-algebraic classification of effective
  theories with enhanced soft limits},''
  \href{http://dx.doi.org/10.1007/JHEP05(2018)076}{{\em JHEP} {\bf 05} (2018)
  076},
\href{http://arxiv.org/abs/1803.05359}{{\tt arXiv:1803.05359 [hep-th]}}.

\bibitem{Roest:2019oiw}
D.~Roest, D.~Stefanyszyn, and P.~Werkman, ``{An Algebraic Classification of
  Exceptional EFTs},''
\href{http://arxiv.org/abs/1903.08222}{{\tt arXiv:1903.08222 [hep-th]}}.

\bibitem{Roest:2019dxy}
D.~Roest, D.~Stefanyszyn, and P.~Werkman, ``{An Algebraic Classification of
  Exceptional EFTs Part II: Supersymmetry},''
\href{http://arxiv.org/abs/1905.05872}{{\tt arXiv:1905.05872 [hep-th]}}.

\bibitem{Bonifacio:2018zex}
J.~Bonifacio, K.~Hinterbichler, A.~Joyce, and R.~A. Rosen, ``{Shift Symmetries
  in (Anti) de Sitter Space},''
  \href{http://dx.doi.org/10.1007/JHEP02(2019)178}{{\em JHEP} {\bf 02} (2019)
  178},
\href{http://arxiv.org/abs/1812.08167}{{\tt arXiv:1812.08167 [hep-th]}}.

\bibitem{deRham:2018svs}
C.~De~Rham, K.~Hinterbichler, and L.~A. Johnson, ``{On the (A)dS Decoupling
  Limits of Massive Gravity},''
  \href{http://dx.doi.org/10.1007/JHEP09(2018)154}{{\em JHEP} {\bf 09} (2018)
  154},
\href{http://arxiv.org/abs/1807.08754}{{\tt arXiv:1807.08754 [hep-th]}}.

\bibitem{Folkerts:2011ev}
S.~Folkerts, A.~Pritzel, and N.~Wintergerst, ``{On ghosts in theories of
  self-interacting massive spin-2 particles},''
\href{http://arxiv.org/abs/1107.3157}{{\tt arXiv:1107.3157 [hep-th]}}.

\bibitem{Hinterbichler:2013eza}
K.~Hinterbichler, ``{Ghost-Free Derivative Interactions for a Massive
  Graviton},'' \href{http://dx.doi.org/10.1007/JHEP10(2013)102}{{\em JHEP} {\bf
  10} (2013)  102},
\href{http://arxiv.org/abs/1305.7227}{{\tt arXiv:1305.7227 [hep-th]}}.

\bibitem{Akagi:2014iea}
S.~Akagi, Y.~Ohara, and S.~Nojiri, ``{New massive spin two model on a curved
  spacetime},'' \href{http://dx.doi.org/10.1103/PhysRevD.90.123013}{{\em Phys.
  Rev.} {\bf D90} (2014) no.~12, 123013},
\href{http://arxiv.org/abs/1410.5553}{{\tt arXiv:1410.5553 [hep-th]}}.

\bibitem{deRham:2010kj}
C.~de~Rham, G.~Gabadadze, and A.~J. Tolley, ``{Resummation of Massive
  Gravity},'' \href{http://dx.doi.org/10.1103/PhysRevLett.106.231101}{{\em
  Phys. Rev. Lett.} {\bf 106} (2011)  231101},
\href{http://arxiv.org/abs/1011.1232}{{\tt arXiv:1011.1232 [hep-th]}}.

\bibitem{deRham:2010ik}
C.~de~Rham and G.~Gabadadze, ``{Generalization of the Fierz-Pauli Action},''
  \href{http://dx.doi.org/10.1103/PhysRevD.82.044020}{{\em Phys. Rev.} {\bf
  D82} (2010)  044020},
\href{http://arxiv.org/abs/1007.0443}{{\tt arXiv:1007.0443 [hep-th]}}.

\bibitem{Fierz:1939ix}
M.~Fierz and W.~Pauli, ``{On relativistic wave equations for particles of
  arbitrary spin in an electromagnetic field},''
\href{http://dx.doi.org/10.1098/rspa.1939.0140}{{\em Proc. Roy. Soc. Lond.}
  {\bf A173} (1939)  211--232}.

\bibitem{Fang:1978se}
J.~Fang and C.~Fronsdal, ``{Elementary Particles in a Curved Space. 5. Massive
  and Massless Spin-2 Fields},''
\href{http://dx.doi.org/10.1007/BF00400165}{{\em Lett. Math. Phys.} {\bf 2}
  (1978)  391--397}.

\bibitem{Dirac:1936fq}
P.~A.~M. Dirac, ``{Wave equations in conformal space},''
\href{http://dx.doi.org/10.2307/1968455}{{\em Annals Math.} {\bf 37} (1936)
  429--442}.

\bibitem{Fronsdal:1978vb}
C.~Fronsdal, ``{Singletons and Massless, Integral Spin Fields on de Sitter
  Space (Elementary Particles in a Curved Space. 7.},''
\href{http://dx.doi.org/10.1103/PhysRevD.20.848}{{\em Phys. Rev.} {\bf D20}
  (1979)  848--856}.

\bibitem{Burrage:2011bt}
C.~Burrage, C.~de~Rham, and L.~Heisenberg, ``{de Sitter Galileon},''
  \href{http://dx.doi.org/10.1088/1475-7516/2011/05/025}{{\em JCAP} {\bf 1105}
  (2011)  025},
\href{http://arxiv.org/abs/1104.0155}{{\tt arXiv:1104.0155 [hep-th]}}.

\bibitem{deRham:2012kf}
C.~de~Rham and S.~Renaux-Petel, ``{Massive Gravity on de Sitter and Unique
  Candidate for Partially Massless Gravity},''
  \href{http://dx.doi.org/10.1088/1475-7516/2013/01/035}{{\em JCAP} {\bf 1301}
  (2013)  035},
\href{http://arxiv.org/abs/1206.3482}{{\tt arXiv:1206.3482 [hep-th]}}.

\bibitem{Gao:2014ula}
X.~Gao, T.~Kobayashi, M.~Yamaguchi, and D.~Yoshida, ``{Covariant Stuckelberg
  analysis of de Rham-Gabadadze-Tolley massive gravity with a general fiducial
  metric},'' \href{http://dx.doi.org/10.1103/PhysRevD.90.124073}{{\em Phys.
  Rev.} {\bf D90} (2014) no.~12, 124073},
\href{http://arxiv.org/abs/1409.3074}{{\tt arXiv:1409.3074 [gr-qc]}}.

\bibitem{Gao:2015xwa}
X.~Gao, ``{Covariant expansion of the gravitational Stuckelberg trick},''
  \href{http://dx.doi.org/10.1103/PhysRevD.91.094001}{{\em Phys. Rev.} {\bf
  D91} (2015) no.~9, 094001},
\href{http://arxiv.org/abs/1502.07691}{{\tt arXiv:1502.07691 [gr-qc]}}.

\bibitem{Hinterbichler:2011tt}
K.~Hinterbichler, ``{Theoretical Aspects of Massive Gravity},''
  \href{http://dx.doi.org/10.1103/RevModPhys.84.671}{{\em Rev. Mod. Phys.} {\bf
  84} (2012)  671--710},
\href{http://arxiv.org/abs/1105.3735}{{\tt arXiv:1105.3735 [hep-th]}}.

\bibitem{deRham:2014zqa}
C.~de~Rham, ``{Massive Gravity},''
  \href{http://dx.doi.org/10.12942/lrr-2014-7}{{\em Living Rev. Rel.} {\bf 17}
  (2014)  7},
\href{http://arxiv.org/abs/1401.4173}{{\tt arXiv:1401.4173 [hep-th]}}.

\bibitem{Hassan:2011tf}
S.~F. Hassan, R.~A. Rosen, and A.~Schmidt-May, ``{Ghost-free Massive Gravity
  with a General Reference Metric},''
  \href{http://dx.doi.org/10.1007/JHEP02(2012)026}{{\em JHEP} {\bf 02} (2012)
  026},
\href{http://arxiv.org/abs/1109.3230}{{\tt arXiv:1109.3230 [hep-th]}}.

\bibitem{Gabadadze:2013ria}
G.~Gabadadze, K.~Hinterbichler, D.~Pirtskhalava, and Y.~Shang, ``{Potential for
  general relativity and its geometry},''
  \href{http://dx.doi.org/10.1103/PhysRevD.88.084003}{{\em Phys. Rev.} {\bf
  D88} (2013) no.~8, 084003},
\href{http://arxiv.org/abs/1307.2245}{{\tt arXiv:1307.2245 [hep-th]}}.

\bibitem{Ondo:2013wka}
N.~A. Ondo and A.~J. Tolley, ``{Complete Decoupling Limit of Ghost-free Massive
  Gravity},'' \href{http://dx.doi.org/10.1007/JHEP11(2013)059}{{\em JHEP} {\bf
  11} (2013)  059},
\href{http://arxiv.org/abs/1307.4769}{{\tt arXiv:1307.4769 [hep-th]}}.

\bibitem{deRham:2010gu}
C.~de~Rham and G.~Gabadadze, ``{Selftuned Massive Spin-2},''
  \href{http://dx.doi.org/10.1016/j.physletb.2010.08.043}{{\em Phys. Lett.}
  {\bf B693} (2010)  334--338},
\href{http://arxiv.org/abs/1006.4367}{{\tt arXiv:1006.4367 [hep-th]}}.

\bibitem{Koyama:2011wx}
K.~Koyama, G.~Niz, and G.~Tasinato, ``{The Self-Accelerating Universe with
  Vectors in Massive Gravity},''
  \href{http://dx.doi.org/10.1007/JHEP12(2011)065}{{\em JHEP} {\bf 12} (2011)
  065},
\href{http://arxiv.org/abs/1110.2618}{{\tt arXiv:1110.2618 [hep-th]}}.

\bibitem{Tasinato:2012ze}
G.~Tasinato, K.~Koyama, and G.~Niz, ``{Vector instabilities and
  self-acceleration in the decoupling limit of massive gravity},''
  \href{http://dx.doi.org/10.1103/PhysRevD.87.064029}{{\em Phys. Rev.} {\bf
  D87} (2013) no.~6, 064029},
\href{http://arxiv.org/abs/1210.3627}{{\tt arXiv:1210.3627 [hep-th]}}.

\bibitem{Yu:2013owa}
S.~Yu, ``{Superluminal Vector in Ghost-free Massive Gravity},''
  \href{http://dx.doi.org/10.1007/JHEP09(2014)019}{{\em JHEP} {\bf 09} (2014)
  019},
\href{http://arxiv.org/abs/1310.6469}{{\tt arXiv:1310.6469 [hep-th]}}.

\bibitem{Deffayet:2005ys}
C.~Deffayet and J.-W. Rombouts, ``{Ghosts, strong coupling and accidental
  symmetries in massive gravity},''
  \href{http://dx.doi.org/10.1103/PhysRevD.72.044003}{{\em Phys. Rev.} {\bf
  D72} (2005)  044003},
\href{http://arxiv.org/abs/gr-qc/0505134}{{\tt arXiv:gr-qc/0505134 [gr-qc]}}.

\bibitem{ArkaniHamed:2002sp}
N.~Arkani-Hamed, H.~Georgi, and M.~D. Schwartz, ``{Effective field theory for
  massive gravitons and gravity in theory space},''
  \href{http://dx.doi.org/10.1016/S0003-4916(03)00068-X}{{\em Annals Phys.}
  {\bf 305} (2003)  96--118},
\href{http://arxiv.org/abs/hep-th/0210184}{{\tt arXiv:hep-th/0210184
  [hep-th]}}.

\bibitem{Bonifacio:2019pfg}
J.~Bonifacio, K.~Hinterbichler, and L.~A. Johnson, ``{Pseudolinear spin-2
  interactions},'' \href{http://dx.doi.org/10.1103/PhysRevD.99.024037}{{\em
  Phys. Rev.} {\bf D99} (2019) no.~2, 024037},
\href{http://arxiv.org/abs/1806.00483}{{\tt arXiv:1806.00483 [hep-th]}}.

\bibitem{Barnich:2001jy}
G.~Barnich and F.~Brandt, ``{Covariant theory of asymptotic symmetries,
  conservation laws and central charges},''
  \href{http://dx.doi.org/10.1016/S0550-3213(02)00251-1}{{\em Nucl. Phys.} {\bf
  B633} (2002)  3--82},
\href{http://arxiv.org/abs/hep-th/0111246}{{\tt arXiv:hep-th/0111246
  [hep-th]}}.

\bibitem{Ivanov:1976zq}
E.~A. Ivanov and V.~I. Ogievetsky, ``{Gauge Theories as Theories of Spontaneous
  Breakdown},''
\href{http://dx.doi.org/10.1007/BF00398486}{{\em Lett. Math. Phys.} {\bf 1}
  (1976)  309--313}.

\bibitem{Goon:2014ika}
G.~Goon, A.~Joyce, and M.~Trodden, ``{Spontaneously Broken Gauge Theories and
  the Coset Construction},''
  \href{http://dx.doi.org/10.1103/PhysRevD.90.025022}{{\em Phys. Rev.} {\bf
  D90} (2014) no.~2, 025022},
\href{http://arxiv.org/abs/1405.5532}{{\tt arXiv:1405.5532 [hep-th]}}.

\bibitem{Goon:2014paa}
G.~Goon, K.~Hinterbichler, A.~Joyce, and M.~Trodden, ``{Einstein Gravity,
  Massive Gravity, Multi-Gravity and Nonlinear Realizations},''
  \href{http://dx.doi.org/10.1007/JHEP07(2015)101}{{\em JHEP} {\bf 07} (2015)
  101},
\href{http://arxiv.org/abs/1412.6098}{{\tt arXiv:1412.6098 [hep-th]}}.

\bibitem{Tasinato:2014eka}
G.~Tasinato, ``{Cosmic Acceleration from Abelian Symmetry Breaking},''
  \href{http://dx.doi.org/10.1007/JHEP04(2014)067}{{\em JHEP} {\bf 04} (2014)
  067},
\href{http://arxiv.org/abs/1402.6450}{{\tt arXiv:1402.6450 [hep-th]}}.

\bibitem{Heisenberg:2014rta}
L.~Heisenberg, ``{Generalization of the Proca Action},''
  \href{http://dx.doi.org/10.1088/1475-7516/2014/05/015}{{\em JCAP} {\bf 1405}
  (2014)  015},
\href{http://arxiv.org/abs/1402.7026}{{\tt arXiv:1402.7026 [hep-th]}}.

\bibitem{Hull:2014bga}
M.~Hull, K.~Koyama, and G.~Tasinato, ``{A Higgs Mechanism for Vector
  Galileons},'' \href{http://dx.doi.org/10.1007/JHEP03(2015)154}{{\em JHEP}
  {\bf 03} (2015)  154},
\href{http://arxiv.org/abs/1408.6871}{{\tt arXiv:1408.6871 [hep-th]}}.

\bibitem{Tasinato:2014mia}
G.~Tasinato, ``{A small cosmological constant from Abelian symmetry
  breaking},'' \href{http://dx.doi.org/10.1088/0264-9381/31/22/225004}{{\em
  Class. Quant. Grav.} {\bf 31} (2014)  225004},
\href{http://arxiv.org/abs/1404.4883}{{\tt arXiv:1404.4883 [hep-th]}}.

\bibitem{Allys:2015sht}
E.~Allys, P.~Peter, and Y.~Rodriguez, ``{Generalized Proca action for an
  Abelian vector field},''
  \href{http://dx.doi.org/10.1088/1475-7516/2016/02/004}{{\em JCAP} {\bf 1602}
  (2016) no.~02, 004},
\href{http://arxiv.org/abs/1511.03101}{{\tt arXiv:1511.03101 [hep-th]}}.

\bibitem{Hull:2015uwa}
M.~Hull, K.~Koyama, and G.~Tasinato, ``{Covariantized vector Galileons},''
  \href{http://dx.doi.org/10.1103/PhysRevD.93.064012}{{\em Phys. Rev.} {\bf
  D93} (2016) no.~6, 064012},
\href{http://arxiv.org/abs/1510.07029}{{\tt arXiv:1510.07029 [hep-th]}}.

\bibitem{Charmchi:2015ggf}
F.~Charmchi, Z.~Haghani, S.~Shahidi, and L.~Shahkarami, ``{One-loop corrections
  to vector Galileon theory},''
  \href{http://dx.doi.org/10.1103/PhysRevD.93.124044}{{\em Phys. Rev.} {\bf
  D93} (2016) no.~12, 124044},
\href{http://arxiv.org/abs/1511.07034}{{\tt arXiv:1511.07034 [hep-th]}}.

\bibitem{Jimenez:2016isa}
J.~Beltran~Jimenez and L.~Heisenberg, ``{Derivative self-interactions for a
  massive vector field},''
  \href{http://dx.doi.org/10.1016/j.physletb.2016.04.017}{{\em Phys. Lett.}
  {\bf B757} (2016)  405--411},
\href{http://arxiv.org/abs/1602.03410}{{\tt arXiv:1602.03410 [hep-th]}}.

\bibitem{Allys:2016jaq}
E.~Allys, J.~P. Beltran~Almeida, P.~Peter, and Y.~Rodriguez, ``{On the 4D
  generalized Proca action for an Abelian vector field},''
  \href{http://dx.doi.org/10.1088/1475-7516/2016/09/026}{{\em JCAP} {\bf 1609}
  (2016) no.~09, 026},
\href{http://arxiv.org/abs/1605.08355}{{\tt arXiv:1605.08355 [hep-th]}}.

\bibitem{Heisenberg:2017mzp}
L.~Heisenberg, ``{Generalised Proca Theories},'' in {\em {Proceedings, 52nd
  Rencontres de Moriond on Gravitation (Moriond Gravitation 2017): La Thuile,
  Italy, March 25-April 1, 2017}}, pp.~233--241.
\newblock 2017.
\newblock
\href{http://arxiv.org/abs/1705.05387}{{\tt arXiv:1705.05387 [hep-th]}}.
\newblock

\bibitem{ErrastiDiez:2019ttn}
V.~Errasti~Díez, B.~Gording, J.~A. Méndez-Zavaleta, and A.~Schmidt-May,
  ``{The complete theory of Maxwell and Proca fields},''
\href{http://arxiv.org/abs/1905.06967}{{\tt arXiv:1905.06967 [hep-th]}}.

\bibitem{ErrastiDiez:2019trb}
V.~Errasti~Diez, B.~Gording, J.~A. Mendez-Zavaleta, and A.~Schmidt-May, ``{The
  Maxwell-Proca theory: definition and construction},''
\href{http://arxiv.org/abs/1905.06968}{{\tt arXiv:1905.06968 [hep-th]}}.

\bibitem{GallegoCadavid:2019zke}
A.~Gallego~Cadavid and Y.~Rodriguez, ``{A systematic procedure to build the
  beyond generalized Proca field theory},''
\href{http://arxiv.org/abs/1905.10664}{{\tt arXiv:1905.10664 [hep-th]}}.

\bibitem{Jimenez:2019hpl}
J.~B. Jim\'enez, C.~de~Rham, and L.~Heisenberg, ``{Generalized Proca and its
  Constraint Algebra},''
\href{http://arxiv.org/abs/1906.04805}{{\tt arXiv:1906.04805 [hep-th]}}.

\bibitem{Joung:2015jza}
E.~Joung and K.~Mkrtchyan, ``{Partially-massless higher-spin algebras and their
  finite-dimensional truncations},''
  \href{http://dx.doi.org/10.1007/JHEP01(2016)003}{{\em JHEP} {\bf 01} (2016)
  003},
\href{http://arxiv.org/abs/1508.07332}{{\tt arXiv:1508.07332 [hep-th]}}.

\bibitem{Bellazzini:2019bzh}
B.~Bellazzini, F.~Riva, J.~Serra, and F.~Sgarlata, ``{Massive Higher Spins:
  Effective Theory and Consistency},''
\href{http://arxiv.org/abs/1903.08664}{{\tt arXiv:1903.08664 [hep-th]}}.

\bibitem{Bonifacio:2017iry}
J.~J. Bonifacio, {\em {Aspects of Massive Spin-2 Effective Field Theories}}.
\newblock PhD thesis, Oxford U., 2017.
\newblock
\url{https://ora.ox.ac.uk/objects/uuid:1e8bde3b-cc1a-4f09-8053-1e05fdd49d49}.
\newblock

\bibitem{Bonifacio:2018vzv}
J.~Bonifacio and K.~Hinterbichler, ``{Bounds on Amplitudes in Effective
  Theories with Massive Spinning Particles},''
  \href{http://dx.doi.org/10.1103/PhysRevD.98.045003}{{\em Phys. Rev.} {\bf
  D98} (2018) no.~4, 045003},
\href{http://arxiv.org/abs/1804.08686}{{\tt arXiv:1804.08686 [hep-th]}}.

\bibitem{Luty:2003vm}
M.~A. Luty, M.~Porrati, and R.~Rattazzi, ``{Strong interactions and stability
  in the DGP model},''
  \href{http://dx.doi.org/10.1088/1126-6708/2003/09/029}{{\em JHEP} {\bf 09}
  (2003)  029},
\href{http://arxiv.org/abs/hep-th/0303116}{{\tt arXiv:hep-th/0303116
  [hep-th]}}.

\bibitem{Nicolis:2004qq}
A.~Nicolis and R.~Rattazzi, ``{Classical and quantum consistency of the DGP
  model},'' \href{http://dx.doi.org/10.1088/1126-6708/2004/06/059}{{\em JHEP}
  {\bf 06} (2004)  059},
\href{http://arxiv.org/abs/hep-th/0404159}{{\tt arXiv:hep-th/0404159
  [hep-th]}}.

\bibitem{Goon:2016ihr}
G.~Goon, K.~Hinterbichler, A.~Joyce, and M.~Trodden, ``{Aspects of Galileon
  Non-Renormalization},'' \href{http://dx.doi.org/10.1007/JHEP11(2016)100}{{\em
  JHEP} {\bf 11} (2016)  100},
\href{http://arxiv.org/abs/1606.02295}{{\tt arXiv:1606.02295 [hep-th]}}.

\bibitem{Folacci:1992xc}
A.~Folacci, ``{BRST quantization of the massless minimally coupled scalar field
  in de Sitter space: Zero modes, euclideanization and quantization},''
  \href{http://dx.doi.org/10.1103/PhysRevD.46.2553}{{\em Phys. Rev.} {\bf D46}
  (1992)  2553--2559},
\href{http://arxiv.org/abs/0911.2064}{{\tt arXiv:0911.2064 [gr-qc]}}.

\bibitem{Miao:2010vs}
S.~P. Miao, N.~C. Tsamis, and R.~P. Woodard, ``{De Sitter Breaking through
  Infrared Divergences},'' \href{http://dx.doi.org/10.1063/1.3448926}{{\em J.
  Math. Phys.} {\bf 51} (2010)  072503},
\href{http://arxiv.org/abs/1002.4037}{{\tt arXiv:1002.4037 [gr-qc]}}.

\bibitem{Garcia-Saenz:2019yok}
S.~Garcia-Saenz, J.~Kang, and R.~Penco, ``{Gauged Galileons},''
\href{http://arxiv.org/abs/1905.05190}{{\tt arXiv:1905.05190 [hep-th]}}.

\bibitem{Elvang:2018dco}
H.~Elvang, M.~Hadjiantonis, C.~R.~T. Jones, and S.~Paranjape, ``{Soft Bootstrap
  and Supersymmetry},'' \href{http://dx.doi.org/10.1007/JHEP01(2019)195}{{\em
  JHEP} {\bf 01} (2019)  195},
\href{http://arxiv.org/abs/1806.06079}{{\tt arXiv:1806.06079 [hep-th]}}.

\end{thebibliography}\endgroup

\end{document}